\documentstyle[preprint,aps]{revtex}
\begin{document}

\newcommand{\s}{\scriptstyle}

\title{On the Exact Solution of Models based on Non-Standard Representations}
\author{J. Gruneberg}
\address{Institut f\"ur Theoretische Physik\\
         Universit\"at zu K\"oln\\
         Z\"ulpicher Stra\ss e 77\\
         50937 K\"oln}
\date{revised version}
\maketitle
\begin{abstract}
The algebraic Bethe ansatz is a powerful method to diagonalize
transfer-matrices of statistical models derived from solutions of 
(graded) Yang Baxter equations, connected to fundamental 
representations of Lie (super-)algebras and their quantum
deformations respectively. It is, however, very difficult to apply 
it to models based on higher dimensional representations of these
algebras in auxiliary space, which are not of fusion type. A systematic 
approach to this problem is presented here. It is illustrated
by the diagonalization of a transfer-matrix of a model based on the 
product of two different four-dimensional representations of 
$U_{q}(\widehat{gl}^{\prime}(2,1;\mathcal{C}))$.\\ 
\end{abstract}
\narrowtext
\section{Introduction}
\label{intro}
The starting point for the construction of (Bethe ansatz) integrable models
is the famous Yang-Baxter equation (YBE) \cite{Yan1,Bax1}.  
\begin{mathletters}
\label{YB-equ}
\begin{eqnarray}
\nonumber
&   & R^{VV^{\prime}}_{12}(u,v)R^{VV^{\prime\prime}}_{13}(u,w)
      R^{V^{\prime}V^{\prime\prime}}_{23}(v,w)\\
\label{YB-equ1}
& = & R^{V^{\prime}V^{\prime\prime}}_{23}(v,w)R^{VV^{\prime\prime}}_{13}(u,w)
      R^{VV^{\prime}}_{12}(u,v)
\end{eqnarray}
$V,V^{\prime}$ and $V^{\prime\prime}$ are three in general different spaces.
The operators $R^{VV^{\prime}}(u)$ act on the direct product 
$V\times V^{\prime}\to V\times V^{\prime}$. Both sides of equation
(\ref{YB-equ1}) act on the three-fold product 
$V\times V^{\prime}\times V^{\prime\prime}$. The lower indices $i,j\in{1,2,3}$
on the $R$-operators denote as usual the two factors in this product on 
which the corresponding $R$-operator acts non-trivially. In general
the so-called \emph{spectral parameters} $u,v$ and $w$ are complex variables.
\\
Up to now, there is no general classification of the solutions to 
(\ref{YB-equ1}). The situation is much better understood, if $V,V^{\prime}$ 
and $V^{\prime\prime}$ are carrier spaces for the representation of a simple 
Lie-algebra or its quantum-deformation. The corresponding theory is mainly due 
to Drinfel'd \cite{Drin}, who also introduced the concept of the
universal $R$-matrix. The existence of the latter guarantees
the existence of $R$-operators as matrices acting on direct products 
of usually, but not always finite dimensional carrier spaces $V$. 
A good account of these developments has been given by Chari and Presley 
\cite{ChPr}. Powerful methods to construct these matrices explicitly were 
developed by Jimbo \cite{Jimb} and many others, see e.g. the book by Ma 
\cite{Ma}. The dependence on only one complex parameter is due to the use of
\emph{evaluation representations} of affine algebras. In this case 
(\ref{YB-equ}) takes the more common difference form
\begin{eqnarray}
\nonumber
&   & R^{VV^{\prime}}_{12}(u-v)R^{VV^{\prime\prime}}_{13}(u)
      R^{V^{\prime}V^{\prime\prime}}_{23}(v)\\
\label{YB-equ2}
& = & R^{V^{\prime}V^{\prime\prime}}_{23}(v)R^{VV^{\prime\prime}}_{13}(u)
      R^{VV^{\prime}}_{12}(u-v)
\end{eqnarray}
\end{mathletters}
The first space $V$ is called \emph{auxiliary space}, the second relabeled to
$V^{(n)}$), in general taken out of some countable set  
$\{V^{(m)}\}_{m=1}^{N}$, a (local) \emph{quantum space}. An
$L$-operator acting on the direct product of these is defined as
\begin{mathletters}
\label{QISM1}
\begin{equation}
\label{L-matrix}
\hat{L}^{V}(n|u)\: :=\: R^{VV^{(n)}}(u,w^{(n)})\:.\\
\end{equation}
It is assumed to act trivially on all other quantum spaces $V^{(m)}$ with 
$m\neq n$. Assuming, that $w^{(n)}$ in (\ref{L-matrix}) just labels 
$V^{(n)}$ and that $u$ is a spectral parameter of difference type as in 
(\ref{YB-equ2}), it is possible to introduce additional 
inhomogeneities $\delta^{(n)}$ into the
\emph{monodromy-matrix}
\begin{equation}
\label{T-matrix}
\hat{T}^{V}(N|u)\::=\:\hat{L}^{V}(N|u-\delta^{(N)})\cdots
                      \cdots\hat{L}^{V}(1|u-\delta^{(1)})\:.
\end{equation}
Here $\delta^{(n)}$ and $w^{(n)}$ will be some complex numbers. 
\begin{equation}
\label{t-matrix}
\hat{\tau}^{V}(N|u)\:=\:\mbox{tr}_{V}\left\{\hat{T}^{V}(N|u)\right\}
\end{equation}
\end{mathletters}
can be viewed as row-to-row \emph{transfer-matrix} of a two dimensional 
(classical) statistical model, with $N$ sites per row, acting on
the (global) quantum space $V^{(N)}\times\cdots\times V^{(1)}$.
If $\delta^{(n)}$ vanishes and $w^{(n)}$ is independent of $n$, 
the \emph{transfer-matrix} (\ref{t-matrix}) is called \emph{homogeneous}. 
In any case integrability of the latter is established via (\ref{YB-equ}), 
written as
\begin{mathletters}
\label{QISM2}
\begin{eqnarray}
\nonumber
&   &
R^{VV^{\prime}}_{12}(u,v)\hat{L}^{V}_{1}(n|u)\hat{L}^{V^{\prime}}_{2}(n|v)\\
\label{CR}
& = & 
\hat{L}^{V^{\prime}}_{2}(n|v)\hat{L}^{V}_{1}(n|u)R^{VV^{\prime}}_{12}(u,v)\:,
\end{eqnarray}
>From that the \emph{fundamental commutation relations} (FCR) are obtained 
immediately:
\begin{eqnarray}
\nonumber
&   & R^{VV^{\prime}}_{12}(u,v)\hat{T}^{V}_{1}(N|u)
                               \hat{T}^{V^{\prime}}_{2}(N|v)\\
\label{FCR}
& = & \hat{T}^{V^{\prime}}_{2}(N|v)
      \hat{T}^{V}_{1}(N|u)R^{VV^{\prime}}_{12}(u,v)
\end{eqnarray}
Provided $R^{VV^{\prime}}_{12}$ is invertible, which is guaranteed for finite
dimensional $V$ and $V^{\prime}$, this yields
\begin{equation}
\label{Integr}
\left[\hat{\tau}^{V}(N|u),\hat{\tau}^{V^{\prime}}(N|v)\right]
\:\stackrel{!}{=}\:0
\end{equation} 
\end{mathletters}
Expanding $\tau^{V^{\prime}}(N|v)$ in $v$ one obtains an infinite 
family of operators commuting with $\hat{\tau}^{V}(N|u)$. The question, 
if this family contains the right number of ``independent'' integrals 
of motion for every finite $N$, is difficult to answer and usually 
taken for granted.\\
The set of equations (\ref{QISM1}) and (\ref{QISM2}) was 
derived by Baxter and can be found together with the original 
references in his excellent book \cite{Baxt}.\\
The notation here is due to Faddeev and coworkers, who 
created a purely algebraic way for diagonalizing $\hat{\tau}^{V}(N|u)$, 
the algebraic Bethe ansatz (ABA). Their quantum inverse scattering 
method (QISM) \cite{QISM} provided the background for Drinfel'd's 
theory \cite{Drin}, but is more general and to the author's opinion 
not fully exploited yet. A good account including original references 
can be found in the book by Korepin et al. \cite{KBI} and the reprint 
collection \cite{Repr}.\\
ABA is a powerful method to construct eigenvectors and eigenvalues of 
$\hat{\tau}^{V}(N|u)$. In some sense it is more systematic than the original
\emph{coordinate Bethe ansatz} \cite{Beth}. In general this is only true, 
if the auxiliary space $V$ is the carrier space of the fundamental  
representation of a Lie (super-)algebra or a deformation of the latter. 
Especially if the auxiliary space $V$ is a higher dimensional carrier 
space of another representation of the same algebra, simplicity is lost 
and ABA becomes cumbersome. Drinfel'd's theory \cite{Drin} suggests, that 
a simple generalization should exist. A systematic approach to this problem 
will be developed in the following.
\section{Models}
\label{models}
In the case of general graded algebras Drinfel'd's constructions \cite{Drin} 
are still not completely understood. However for simple (affine) Lie 
superalgebras and their quantum deformations a proper algebraic construction 
has been given by Yamane recently \cite{Yama}. Also QISM and ABA are not very 
sensitive to grading and the graded version of the YBE has been established 
by Kulish and Sklyanin long ago \cite{KuSk}.\\
The $R$-matrices, which will be used as concrete examples, are related to   
the ``quantum universal enveloping superalgebra''
$U_{q}(\widehat{gl}^{\prime}(2,1|\mathcal{C}))$. No use will be made of any 
peculiar features of this symmetry. The interested reader is referred to the 
book by Cornwell \cite{Corn} on Lie superalgebras, from which the notation 
is borrowed, the book by Kac \cite{Kac} for more details on affinization
and to the paper \cite{Yama} for the proper construction of the
$q$-deformed universal enveloping superalgebra.\\
The carrier space $V_{3}$ of the fundamental representation of 
$U_{q}(gl(2,1|{\mathcal C})$ is complex and three-dimensional. Basis and 
cobasis will be denoted by
\begin{mathletters}
\begin{equation}
\label{basis}
|i>\:,\quad <j|i>\:=\:\delta_{ij}\quad\mbox{for}\quad i,j\:=\:1,2,3
\end{equation}
A basis of the complex carrier space $V_{4}$ of the four-dimensional
representation will be denoted similarly. These representations are 
${\mathcal Z}_{2}$-graded: To each basis-vector $\left|i\right>$ a number 
$p(i)\in\{0,1\}$ is assigned, i.e.
\begin{equation}
\label{grading1}
p(1)\:=\:p(2)\:=\:0\quad,\qquad p(3)\:=\:1
\end{equation}
for $V_{3}$ and analogously 
\begin{equation}
\label{grading2}
p(1)\:=\:p(2)\:=\:0\quad,\qquad p(3)\:=\:p(4)\:=\:1
\end{equation}
for $V_{4}$. Local basis-vectors are divided into even (bosonic, $p=0$) 
and odd (fermionic, $p=1$) ones. Local operators acting in $V_{3}$ 
or $V_{4}$ etc. are expressed in the natural basis
\begin{equation}
\label{Weyl-b}
e_{ij}\:=\:\left|i\right>\left<j\right|
\end{equation}
If the corresponding space is a (local) quantum space, it will be denoted with
a hat, e.g. $\hat{e}_{ij}$ for clarity. These operators act trivially 
on all other (local) quantum spaces. A grading is assigned to this
basis according to
\begin{equation}
\label{m-grad}
p(e_{ij})\:=\:\left[p(i)\,+\,p(j)\right]\:\mbox{mod}\:2
\end{equation}
It is possible to extend these definitions of grading naturally to those 
vectors $\left|\psi\right>$ and operators $\hat{a}$, which are
homogeneous with respect to the grading.\\
It is convenient to expand operators as well as vectors in the natural 
(tensor) product basis, which is ordered according to 
$V^{(N)}\times\cdots\times V^{(1)}$, see (\ref{T-matrix}). 
Grading imposes signs on products of homogeneous operators, i.e.:
\begin{equation}
\label{g-mult1}
(\hat{a}\otimes\hat{b}) 
(\hat{c}\otimes\hat{d})\:=\:(-1)^{p(\hat{b})p(\hat{c})}
(\hat{a}\hat{c})\otimes(\hat{b}\hat{d})
\end{equation}
or on the action of homogeneous operators on homogeneous vectors, i.e.
\begin{eqnarray}
\nonumber
&   & (\hat{a}\otimes\hat{b}) 
      (\left|\psi\right>\otimes\left|\varphi\right>)\\
\label{g-mult2}
& = & (-1)^{p(\hat{b})p(\left|\psi\right>)} 
      (\hat{a}\left|\psi\right>)\otimes(\hat{b}\left|\varphi\right>)
\end{eqnarray}
The only other effect of grading is, that $tr_{V}$ in (\ref{t-matrix})
has to be interpreted as \emph{supertrace}:
\begin{equation}
\label{s-tr}
\mbox{tr}_{V}\left\{\hat{T}^{V}(N|u)\right\}\:=\:\sum_{i}(-1)^{p(i)}
\langle i|\hat{T}^{V}(N|u)|i\rangle\:.
\end{equation}
\end{mathletters}
Kulish and Sklyanin found \cite{KuSk}, that additional signs, which 
appear in an explicit representation of the YBE (\ref{YB-equ}) due to 
grading can be absorbed into a redefinition of matrix elements, so 
that every solution of the graded YBE is equivalent to a solution
of the conventional one.\\
The four dimensional representation can be characterized by a set of
complex parameters, symbolically denoted by 
\begin{mathletters}
\begin{equation}
\label{V4para}
V_{4}\:\approx\:\left\{C,\kappa,\kappa^{\ast},\mu,\mu^{\ast}\right\}\:.
\end{equation}
This is a peculiarity of Lie superalgebras \cite{Corn}, which 
is conserved under quantum deformation; $\kappa,\kappa^{\ast}$ and 
$\mu,\mu^{\ast}$ are not necessarily complex conjugate to each other,
but related to $C$ by
\begin{equation}
\label{C-cond}
\kappa\kappa^{\ast}\:=\:[C]_{q}\quad,\qquad\mu\mu^{\ast}\:=\:[C+1]_{q}\:,
\end{equation}
where $q$ is the deformation parameter, 
\begin{equation}
q\::=\:e^{2\eta}\:,
\end{equation}
and $q$-brackets are defined as usual by
\begin{equation}
\label{$q$-bracket}
[C]_{q}\::=\:\frac{q^{C}\,-\,q^{-C}}{q\,-\,q^{-1}}\:=\:
             \frac{\sinh(2\eta C)}{\sinh(2\eta)} 
\end{equation}
\end{mathletters}
Different choices of $\kappa,\kappa^{\ast},\mu,\mu^{\ast}$
can be related to each other by a similarity transformation of the algebra, 
which conserves grading, but is not unitary in general. That makes it 
convenient to keep these parameters. Note that the representation 
$V_{4}$ can be deformed continuously into $V^{\prime}_{4}$, which is 
characterized by a set of primed parameters also connected by (\ref{C-cond}).
\\
A well-known solution of (\ref{YB-equ2}) with 
$V=V^{\prime}=V^{\prime\prime}=V_{3}$ is
\begin{eqnarray}
\label{RV3V3}
R^{V_{3}V_{3}}(u) & = & e_{11}\otimes\hat{e}_{11} + e_{22}\otimes\hat{e}_{22}
                        - d(u)\,e_{33}\otimes\hat{e}_{33}\\
\nonumber
                  & + & c(u)\left[e_{11}(\hat{e}_{22}+\hat{e}_{33}) 
                                 +e_{22}(\hat{e}_{11}+\hat{e}_{33})\right]\\
\nonumber
                  & + & a(u)e_{21}\otimes\hat{e}_{12}+
                        b(u)e_{12}\otimes\hat{e}_{21}\\
\nonumber
                  & + & a(u)\left[e_{31}\otimes\hat{e}_{13}+
                                  e_{32}\otimes\hat{e}_{23}\right]\\
\nonumber
                  & - & b(u)\left[e_{13}\otimes\hat{e}_{31}+
                                  e_{23}\otimes\hat{e}_{32}\right]
\end{eqnarray}
with coefficients
\begin{eqnarray*}
a(u) & := & \frac{\sinh(2\eta)}{\sinh(2\eta+u)}
            \left[\cosh(u)+\sinh(u)\right]\\
b(u) & := & \frac{\sinh(2\eta)}{\sinh(2\eta+u)}
            \left[\cosh(u)-\sinh(u)\right]\\
c(u) & := & \frac{\sinh(u)}{\sinh(2\eta+u)}\\
d(u) & := & \frac{\sinh(2\eta-u)}{\sinh(2\eta+u)}
\end{eqnarray*}
To the authors knowledge it appeared first in a different notation in
the work of Perk and Schultz \cite{PeSc}. It is the standard $q$-deformation 
of the $Y(gl(2,1|{\mathcal C}))$-symmetric $R$-matrix given by Kulish and 
Sklyanin \cite{KuSk}.\\
Kulish and Sklyanin wrote down the $Y(gl(m,n|{\mathcal C}))$-symmetric 
$R$-matrix for arbitrary positive integers $m$ and $n$. Its generalization 
to the $U_{q}(\widehat{gl}^{\prime}(m,n|{\mathcal C}))$-symmetric case
can also be taken from the paper by Perk and Schultz \cite{PeSc}. It is 
a simple generalization of (\ref{RV3V3}).\\
The $R$-matrix (\ref{RV3V3}) is related to the following 
$U_{q}(\widehat{gl}^{\prime}(2,1|{\mathcal C}))$-symmetric $R$-matrix
\begin{eqnarray}
\label{RV3V4}
&   & R^{V_{3}V_{4}}(u)\\
\nonumber
& = & \rho(u)\left[e_{11}\otimes(\hat{e}_{11} + \hat{e}_{33})
      \,+\,e_{22}\otimes(\hat{e}_{11} + \hat{e}_{44})\right]\\
\nonumber
& + & \alpha_{0}(u)\left[e_{11}\otimes(\hat{e}_{22} + \hat{e}_{44})
      \,+\,e_{22}\otimes(\hat{e}_{22} + \hat{e}_{33})\right]\\
\nonumber
& + & e_{33}\otimes\left[\beta_{0}(u)\hat{e}_{11} - \hat{e}_{22} + 
                         \gamma_{0}(u)(\hat{e}_{33} + \hat{e}_{44})\right]\\
\nonumber
& + & \delta_{1}(u)e_{12}\otimes \hat{e}_{43} + 
      \delta_{2}(u)e_{21}\otimes \hat{e}_{34}\\
\nonumber
& - & \varepsilon_{1}(u)\left[e_{13}\otimes\hat{e}_{23} + 
                              e_{23}\otimes\hat{e}_{24} \right]\\
\nonumber
& + & \varepsilon_{2}(u)\left[e_{31}\otimes\hat{e}_{32} +
                              e_{32}\otimes\hat{e}_{42} \right]\\
\nonumber
& + & \delta_{1}(u)e_{12}\otimes\hat{e}_{43} +
      \delta_{2}(u)e_{21}\otimes\hat{e}_{34}\\
\nonumber
& - & \zeta_{1}(u)\left[e_{13}\otimes\hat{e}_{41} -
                  q^{-1}e_{23}\otimes\hat{e}_{31}\right]\\
\nonumber
& + & \zeta_{2}(u)\left[e_{31}\otimes\hat{e}_{14} -
                     q\,e_{32}\otimes\hat{e}_{13}\right]
\end{eqnarray}
with coefficients (\ref{V3V4-coeff}), listed in appendix \ref{AppA}, in the
sense, that it fulfills the YBE (\ref{YB-equ2}) with 
$V=V^{\prime}=V_{3}$ and $V^{\prime\prime}=V_{4}$:
\begin{eqnarray}
\nonumber
&   & R^{V_{3}V_{3}}_{12}(u-v)R^{V_{3}V_{4}}_{13}(u)
      R^{V_{3}V_{4}}_{23}(v)\\
\label{YB1}
& = & R^{V_{3}V_{4}}_{23}(v)R^{V_{3}V_{4}}_{13}(u)
      R^{V_{3}V_{3}}_{12}(u-v)
\end{eqnarray}
The construction of the $R$-matrix (\ref{RV3V4}), and the proof of 
(\ref{YB1}) is standard (see e.g.\cite{Ma,Gru1}).\\
>From (\ref{RV3V4}) a \emph{transfer-matrix} $\tau^{V_{3}}(N|u)$ is defined by
(\ref{QISM1}). It is sufficient to consider the homogeneous case, 
i.e. $\delta^{(n)}=0$ and $V^{(n)}=V_{4}$ for all $n$ in (\ref{L-matrix}). 
Integrability follows from (\ref{Integr}). It is easily tractable by ABA, 
which will be demonstrated in the next section.\\ 
Another $U_{q}(\widehat{gl}^{\prime}(2,1|\mathcal{C}))$-symmetric $R$-matrix 
acting on the direct product of two different four dimensional representations,
characterized by the corresponding parameter sets (\ref{V4para}), is given by
\begin{eqnarray}
\label{RV4V4}
&   & R^{V_{4}V^{\prime}_{4}}(u)\\
\nonumber 
& = & f(u)e_{11}\otimes\hat{e}_{11}+g(u)e_{22}\otimes\hat{e}_{22}
      - e_{33}\otimes\hat{e}_{33}  -e_{44}\otimes\hat{e}_{44}\\
\nonumber
& + & r_{5}e_{22}\otimes\hat{e}_{11}+r^{\prime}_{5}e_{11}\otimes\hat{e}_{22}
      -r_{10}(e_{33}\otimes\hat{e}_{44}-e_{44}\otimes\hat{e}_{33})\\
\nonumber
& - & r_{7}(e_{33}+e_{44})\otimes\hat{e}_{11} 
      - r^{\prime}_{7}e_{11}\otimes(\hat{e}_{33}+\hat{e}_{44})\\
\nonumber
& - & r_{9}(e_{33}+e_{44})\otimes\hat{e}_{22} 
      - r^{\prime}_{9}e_{22}\otimes(\hat{e}_{33}+\hat{e}_{44})\\
\nonumber
& + & r_{1}e_{21}\otimes\hat{e}_{12} +r^{\prime}_{1}e_{12}\otimes\hat{e}_{21}
      -r_{4}e_{43}\otimes\hat{e}_{34}-r^{\prime}_{4}e_{34}\otimes\hat{e}_{43}\\
\nonumber
& + & r_{2}(e_{31}\otimes\hat{e}_{13}+e_{41}\otimes\hat{e}_{14})
      -r^{\prime}_{2}(e_{13}\otimes\hat{e}_{31}+e_{14}\otimes\hat{e}_{41})\\
\nonumber
& + & r_{3}(e_{32}\otimes\hat{e}_{23}+e_{42}\otimes\hat{e}_{24})
      -r^{\prime}_{3}(e_{23}\otimes\hat{e}_{32}+e_{24}\otimes\hat{e}_{42})\\
\nonumber
& - & r_{6}(e_{24}\otimes\hat{e}_{13}-q^{-1}e_{23}\otimes\hat{e}_{14})\\
\nonumber
& + & r^{\prime}_{8}(e_{42}\otimes\hat{e}_{31}-q\,e_{32}\otimes\hat{e}_{41})\\
\nonumber
& + & r^{\prime}_{6}(e_{13}\otimes\hat{e}_{24}-q\,e_{14}\otimes\hat{e}_{23})\\
\nonumber
& - & r_{8}(e_{31}\otimes\hat{e}_{42}-q^{-1}e_{41}\otimes\hat{e}_{32})
\end{eqnarray}
The coefficients are again listed in appendix \ref{AppA}. The construction 
of this $R$-matrix, and a proof of (\ref{YB-equ2}),
\begin{mathletters} 
\begin{eqnarray}
\nonumber
&   & R^{V_{4}V^{\prime}_{4}}_{12}(u-v)R^{V_{4}V^{\prime\prime}_{4}}_{13}(u)
      R^{V^{\prime}_{4}V^{\prime\prime}_{4}}_{23}(v)\\
\label{YB2}
& = & R^{V^{\prime}_{4}V^{\prime\prime}_{4}}_{23}(v)
      R^{V_{4}V^{\prime\prime}_{4}}_{13}(u)R^{V_{4}V^{\prime}_{4}}_{12}(u-v)
\end{eqnarray}
or
\begin{eqnarray}
\nonumber
&   & R^{V_{3}V_{4}}_{12}(u-v)R^{V_{3}V^{\prime}_{4}}_{13}(u)
      R^{V_{4}V^{\prime}_{4}}_{23}(v)\\
\label{YB3}
& = & R^{V_{4}V^{\prime}_{4}}_{23}(v)
      R^{V_{3}V^{\prime}_{4}}_{13}(u)R^{V_{3}V_{4}}_{12}(u-v)
\end{eqnarray}
\end{mathletters}
with $R^{V_{3}V_{4}}$ from (\ref{RV3V4}) will be given elsewhere \cite{Gru1}. 
A special case ($V^{\prime}_{4}=V_{4}$), leading to considerable 
simplifications, has been constructed explicitly by Gould et al. 
\cite{GHLZ}.\\
One may fix $u$ and $v$ in (\ref{YB2}) and regard $C,C^{\prime}$ and 
$C^{\prime\prime}$ instead as spectral parameters in order to satisfy 
the general form (\ref{YB-equ1}) of the YBE.\\
>From (\ref{RV4V4}) the \emph{transfer-matrix} $\tau^{V_{4}}(N|u)$ is defined by
(\ref{QISM1}). It is again sufficient to consider the homogeneous case 
$\delta^{(n)}=0$ and $V^{(n)}=V^{\prime}_{4}$ for all $n$ in (\ref{T-matrix}). 
Integrability follows from (\ref{Integr}) with the choice 
between $\tau^{V_{4}}(N|v)$ and $\tau^{V_{3}}(N|v)$ as generating 
functionals for ``integrals of motion''.\\
Here ABA is not straight-forward. This model requires a new strategy 
in order to obtain equations for all eigenvalues $\tau^{V_{4}}(N|u)$.
\section{Algebraic Bethe ansatz}
\label{ABA}
The original recipe for ABA is simple \cite{QISM}:
\begin{enumerate}
\item Determine a \emph{vacuum state}, preferably a highest or lowest weight
      state of the underlying group structure, if available, 
      which tridiagonalizes 
      $\hat{L}^{V}(n|u)$ locally, and extend it via the product structure 
      (\ref{T-matrix}) to a global vacuum, tridiagonalizing 
      $\hat{T}^{V}(N|u)$. 
\item Take the off-diagonal elements of $\hat{T}^{V}(N|u)$, not
      annihilating the global vacuum, as \emph{creation-operators} and use 
      the associative algebra defined by the FCR (\ref{FCR}), to generate
      eigenvectors to all eigenvalues of $\hat{\tau}^{V}(N|u)$
      (\ref{t-matrix}). Equations determining the latter are also
      derived from the algebra. 
\end{enumerate}
The first point is more or less a precondition for the applicability
of ABA; the second is crucial: Only if $V$ is a carrier space of the
fundamental representation of a possibly deformed and graded Lie
algebra, the choice of \emph{creation-operators} is obvious.\\
$\hat{\tau}^{V_{3}}(N|u)$ and $\hat{\tau}^{V_{4}}(N|u)$, as defined in 
the previous section, are sufficiently complex to illustrate the
general situation.\\
Since the auxiliary space is graded, it is useful to transform the
matrix-elements of $\hat{L}^{V_{3}}(n|u)$ (\ref{L-matrix}) in the 
$V_{3}$ basis according to
\begin{equation}
\label{rescal}
\left[\hat{L}^{V_{3}}(n|u)\right]^{(V_{3})}_{ij}\:\to\:
(-1)^{p(j)[p(i)+p(j)]}\left[\hat{L}^{V_{3}}(n|u)\right]^{(V_{3})}_{ij}
\end{equation}   
This absorbs just a troublesome minus sign from the commutation of 
$\left|3\right>^{V_{3}}$ with $[\hat{L}^{V_{3}}(n|u)]_{13}$ and
$[\hat{L}^{V_{3}}(n|u)]_{23}$. All four local basis vectors of 
$V^{(n)}_{4}$ (\ref{basis}) are suitable as (local) vacuum,
preferably
\begin{equation}
\label{l-vac}
\Omega^{(n)}\::=\:|2>^{(n)}
\end{equation}
$\Omega^{(n)}$ is a lowest weight state of the representation of 
$U_{q}(gl(2,1|{\mathcal C}))$ on $V^{(n)}_{4}$ and its 
equivalent was used by Kulish and Reshetikhin to treat the non-graded
$Y(gl(3|{\mathcal C}))$-symmetric case \cite{KuRe}. Their calculation
was generalized to the fundamental representation of 
$U_{q}(\widehat{gl}^{\prime}(m,n))$ by Schultz \cite{Schu}.
\begin{equation}
\label{l-vac-act1}
     \hat{L}^{V_{3}}(n|u)\Omega^{(n)}
\:=\:\left(\begin{array}{ccc}
           \omega^{(n)}_{1}(u) & 0 & 0\\
           0 & \omega^{(n)}_{2}(u) & 0\\
           \ast & \ast & \omega^{(n)}_{3}(u)
           \end{array}\right)\Omega^{(n)}
\end{equation}
with $\ast$ denoting non-zero entries. The vacuum-eigenvalues of the 
diagonal elements are given by.
\begin{eqnarray}
\nonumber
\omega^{(n)}_{1}(u) & = & \frac{\sinh(\eta C +u)}{\sinh(\eta(C+2)-u)}\\
\label{l-vac-ev}
\omega^{(n)}_{2}(u) & = & \omega^{(n)}_{1}(u)\\
\nonumber
\omega^{(n)}_{3}(u) & = & -1
\end{eqnarray}
The index $(n)$ will be omitted, due to homogeneity. Immediately from 
(\ref{l-vac-act1}), (\ref{T-matrix}) and the definition
\begin{equation}
\label{g-vac}
\left|0\right>_{N} \:=\:\Omega^{(N)}\otimes\Omega^{(N-1)}\otimes\cdots
                        \otimes\Omega^{(1)}
\end{equation}
of the (global) vacuum $\left|0\right>_{N}$ follows
\begin{eqnarray}
\nonumber
&   & \hat{T}^{V_{3}}(N|u)\,\left|0\right>_{N}\\
\label{g-vac-act1}
& = & \left(\begin{array}{ccc}
            [\omega_{1}(u)]^{N} & 0 & 0\\
            0 & [\omega_{1}(u)]^{N} & 0\\
            \hat{C}_{1}(u) & \hat{C}_{2}(u) & (-1)^{N}
            \end{array}\right)\left|0\right>_{N}
\end{eqnarray}
where
\begin{equation}
\label{C-op} 
\hat{C}_{i}(u)\::=\:[\hat{T}^{V_{3}}(N|u)]_{3i}
                    \quad\mbox{for}\quad i\,=\,1,2 
\end{equation}
will later serve as \emph{creation-operators}.\\
ABA step 1 is finished: From (\ref{g-vac-act1}),(\ref{t-matrix}) 
and (\ref{s-tr}) follows the vacuum-eigenvalue of $\tau^{V_{3}}(N|u)$: 
\begin{equation}
\label{V3-vac-ev}
\Lambda^{V_{3}}_{N}(u)\:=\:2[\omega_{1}(u)]^{N}\:-\:(-1)^{N}
\end{equation}
As mentioned before, Kulish and Reshetikhin solved a model built from 
the fundamental representation of $Y(gl(3|{\mathcal C}))$, whose 
$R$-matrix differs from the $\eta\to 0$-limit of (\ref{RV3V3}) only in 
minor details. The FCR (\ref{FCR}) derived from (\ref{YB1}):
\begin{eqnarray}
\nonumber
&   & R^{V_{3}V_{3}}_{12}(u-v)\hat{T}^{V_{3}}_{1}(N|u)
                              \hat{T}^{V_{3}}_{2}(N|v)\\
\label{FCR1}
& = & \hat{T}^{V_{3}}_{2}(N|v)
      \hat{T}^{V_{3}}_{1}(N|u)R^{V_{3}V_{3}}_{12}(u-v)
\end{eqnarray}
are almost identical to the ones in \cite{KuRe}: Trigonometric functions in 
(\ref{RV3V3}) do not show up, if appropriate abbreviations are used.
Apart from a few signs due to grading, which was also realized in
\cite{KuRe}, the formal algebra defined by (\ref{FCR1}) becomes exactly the 
same.\\
Of course it is possible to write down equations for eigenvectors and 
eigenvalues immediately, using the result of \cite{KuRe}. Again apart from a 
few signs, just the vacuum eigenvalues have to be replaced by 
(\ref{l-vac-ev}). This is a well-know feature of ABA.\\
However some more details will be needed, in order to tackle the more 
complicated problem of diagonalizing $\hat{\tau}^{V_{4}}(N|u)$ in the 
following section:\\
The (nested, see below) algebraic Bethe ansatz for (right) eigenvectors of 
$\hat{\tau}^{V_{3}}(N|u)$ is \cite{KuRe}
\begin{eqnarray}
\nonumber
&   & |\lambda_{1},\ldots,\lambda_{M}|F>\\
\label{NABA-es}
& = & F^{a_{1},\ldots,a_{M}}\hat{C}_{a_{1}}(\lambda_{1})\cdots
      \hat{C}_{a_{M}}(\lambda_{M})\left|0\right>_{N}\:,
\end{eqnarray}
where $\{\lambda_{1},\ldots,\lambda_{M}\}$ is some set of yet unknown 
parameters and $F^{a_{1},\ldots,a_{M}}$ are some coefficients, 
yet undetermined. Summation over repeated $a_{i}=1,2$ with $i=1,\ldots,M$ 
is implied. From (\ref{FCR1}) follows immediately
\begin{mathletters}
\label{FCR2}
\begin{eqnarray}
\nonumber
      \hat{T}_{33}(u)\hat{C}_{i}(v) 
& = & \frac{1}{c(u-v)}\,\hat{C}_{i}(v)\hat{T}_{33}(u)\\
\label{FCR2a}
& + & \frac{a(v-u)}{c(v-u)}\,\hat{C}_{i}(u)\hat{T}_{33}(v)\\ 
\nonumber
      \hat{T}_{ij}(u)\hat{C}_{k}(v)
\nonumber
& = & \frac{1}{c(u-v)}\sum_{l,m=1}^{2}r_{lm,jk}(u-v)\hat{C}_{m}(v)
      \hat{T}_{il}(u)\\
\label{FCR2b}
& - & \frac{b(u-v)}{c(u-v)}\,\hat{C}_{j}(u)\hat{T}_{ik}(v)\\
\nonumber
      \hat{C}_{i}(u)\hat{C}_{j}(v)
& = & \frac{1}{d(u-v)}\sum_{k,l=1}^{2}r_{kl,ij}(u-v)\hat{C}_{l}(v)
      \hat{C}_{k}(u)\\
\label{FCR2c}
&   &
\end{eqnarray}
\end{mathletters}
with $i,j,k\in\{1,2\}$. $a(u),b(u),c(u)$ and $d(u)$ originate from 
(\ref{RV3V3}). For brevity $[\hat{T}^{V_{3}}(N|u)]^{V_{3}}_{ij}$ 
has been denoted by  $\hat{T}_{ij}(u)$. In the present case 
$r_{ik,jl}(u)$ are elements of the non-graded
$U_{q}(\widehat{gl}^{\prime}(2|{\mathcal C}))$-symmetric $R$-matrix, 
\begin{eqnarray}
\label{R6V}
R^{V_{2}V_{2}}(u) & = & \sum_{i,j,k,l=1}^{2}
           r_{ik,jl}e_{ij}\otimes\hat{e}_{kl}\\ 
\nonumber
     & = & e_{11}\otimes\hat{e}_{11}\:+\:e_{22}\otimes\hat{e}_{22}\\
\nonumber
     & + & c(u)\left[e_{11}\otimes\hat{e}_{22}
           \:+\:e_{22}\otimes\hat{e}_{11}\right]\\
\nonumber
     & + & a(u)e_{21}\otimes\hat{e}_{12}\:+\:b(u)e_{12}\otimes\hat{e}_{21}
\end{eqnarray}
which acts on the direct product of two two-dimensional, purely even
subspaces $V_{2}$ of $V_{3}$, spanned by $|1>$ and $|2>$ from (\ref{basis}).
It is crucial to realize the appearance of $R^{V_{2}V_{2}}(u)$ as a proper 
submatrix in $R^{V_{3}V_{3}}(u)$ (\ref{RV3V3}), because it defines a simpler 
BA-solvable model. Nested algebraic Bethe ansatz (NABA) is typical for 
models, based on fundamental representations of dimension larger than 2.\\
It was preceded by the ingenious, but complicated nested \emph{coordinate} 
Bethe ansatz, invented by Gaudin \cite{Gaud} and Yang \cite{Yan1} 
independently. Their method was applied to the fundamental representation of 
the $Y(gl(m,n|{\mathcal C}))$-symmetric problem by Lai \cite{Lai} and 
Sutherland \cite{Suth}. The formal algebraic formulation of the method is 
apparently due to Takhtajan \cite{Tak1}.\\
The transfer-matrix $\hat{\tau}^{V_{3}}(N|u)$ applied to the Bethe ansatz 
eigenvector (\ref{NABA-es}) should yield 
\begin{eqnarray}
\nonumber 
&   & \hat{\tau}^{V_{3}}(N|u)\,|\lambda_{1},\ldots,\lambda_{M}|F>\\
\label{ev-equ}
& = & \Lambda^{V_{3}}(N|u)\,|\lambda_{1},\ldots,\lambda_{M}|F>
\end{eqnarray}
Leaving some technical details for appendix \ref{AppB}, it turns out,  
that this is true, iff the coefficients $F$ in (\ref{NABA-es}) fulfill
``6-vertex-type'' eigenvalue equations \cite{KuRe}: 
\begin{eqnarray}
\nonumber
&   & \left[\hat{\tau}^{V_{2}}(M|\lambda_{k})
      \right]^{a_{1},\ldots,a_{m}}_{b_{1},\ldots,b_{m}}
      F^{b_{1},\ldots,b_{m}}\\
\label{6V-ev-equ}
& = & \frac{1}{[-\omega_{1}(\lambda_{k})]^{N}}F^{a_{1},\ldots,a_{m}}
\end{eqnarray}
for $k=1,\ldots,M$, of course solvable by ABA \cite{QISM}. This is the
second \emph{nested} Bethe ansatz. $\hat{\tau}^{V_{2}}(M|u)$ is an 
inhomogeneous \emph{transfer-matrix} obtained according to (\ref{QISM1}) 
with $\delta^{(n)}=\gamma_{n}$ from (\ref{R6V}). The eigenvalue of 
$\tau^{V_{2}}(M|u)$ corresponding to the BA-eigenvector $F$ is given by 
\begin{eqnarray}
\nonumber
&   & \Lambda^{V_{2}}_{M}(u;\mu_{1},\ldots,\mu_{m})\\
\label{6V-ev}
& = & \left(\prod_{n=1}^{M}c(u-\lambda_{n})\right)
\left(\prod_{\alpha =1}^{m}\frac{1}{c(u-\mu_{\alpha})}\right)\\
\nonumber
& + & \qquad\qquad\qquad\qquad\left(\prod_{\alpha=1}^{m}
      \frac{1}{c(\mu_{\alpha}-u)}
      \right)
\end{eqnarray}
with \emph{rapidities} $\mu_{\alpha}\:(\alpha=1,\ldots,m)$, determined by 
the BA-equations
\begin{mathletters}
\label{BA-equ1}
\begin{equation}
\label{BA-equ1a}
     \prod_{n=1}^{M}c(\mu_{\alpha}-\lambda_{n})
\:=\:\prod_{\stackrel{\s \beta =1}{\s \beta\neq\alpha}}^{m}
     \frac{c(\mu_{\alpha}-\mu_{\beta})}{c(\mu_{\beta}-\mu_{\alpha})}
\end{equation}
for $\alpha=1,\ldots,m$. These and expressions for the actual BA-vectors
$F$ also depending on $\mu_{1},\ldots,\mu_{m}$, may be found in the 
literature \cite{QISM}.\\
Using (\ref{6V-ev}) the eigenvalue condition (\ref{ev-equ}) reads
\begin{equation}
\label{BA-equ1b}
     [-\omega_{1}(\lambda_{k})]^{N}
\:=\:\prod_{\alpha=1}^{m}c(\mu_{\alpha}-\lambda_{k})
\end{equation}
\end{mathletters}
for $k=1,\ldots,M$, which is the second set of BA-equations, determining 
$\lambda_{1},\ldots,\lambda_{M}$. Collecting the \emph{wanted terms}
in (\ref{FCR3}) the eigenvalue of $\hat{\tau}^{V_{3}}(N|u)$
corresponding to the NABA-eigenvector (\ref{NABA-es}) follows immediately:
\begin{eqnarray}
\label{V3-ev1}
&   & \Lambda^{V_{3}}_{N}(u;\lambda_{1},\ldots,\lambda_{M}
                           |\mu_{1},\ldots,\mu_{m})\\
\nonumber
& = & \left(\prod_{i=1}^{M}\frac{1}{c(u-\lambda_{i})}\right)\\
\nonumber
& \times & \left\{[\omega_{1}(u)]^{N}
                  \Lambda^{V_{2}}_{M}(u;\mu_{1},\ldots,\mu_{m})
                  \,-\,(-1)^{N}\right\}
\end{eqnarray}
According to Baxter \cite{Baxt} BA-equations guarantee analyticity
of all eigenvalues in $u$. Here a $q$-deformed, graded version of the 
$R$-matrix (\ref{RV3V3}) has been used and the $\hat{C}_{i}$-operators 
act on a different quantum space, i.e. $V_{4}$ instead of $V_{3}$. 
However not knowing about \cite{Schu}, the whole calculation has been borrowed 
from \cite{KuRe}. A highest weight state, i.e. $\left|1\right>$ instead of 
$\left|2\right>$ in (\ref{l-vac}) and (\ref{g-vac}), could have been used 
as vacuum, but this leads to a very similar calculation.\\
The result (\ref{V3-ev1}) is new, but it differs just by the vacuum 
eigenvalues (\ref{l-vac-ev}) and signs from the well-known one in \cite{KuRe}. 
It is also complete. This is not true for the set of eigenvectors 
(\ref{NABA-es}). However the missing ones may be produced using the lowest 
weight property of the ABA-vectors with respect to the group action on quantum 
space, which can be proved by standard-methods \cite{QISM}.\\
These are well-known and beautiful features of Bethe ansatz 
solvable systems. Also the equations for the inhomogeneous model with 
$w^{(n)}=C^{(n)}$ in (\ref{L-matrix}) can be written down immediately 
using an argument due to Baxter \cite{Baxt}:
\begin{mathletters}
\label{V3-sol}
\begin{eqnarray}
\label{V3-ev2}
&   & \Lambda^{V_{3}}_{N}(u;\lambda_{1},\ldots,\lambda_{M}|
                            \mu_{1},\ldots,\mu_{m})\\
\nonumber
& = & \prod_{n=1}^{N}\left(\frac{\sinh(\eta C^{(n)}+u-\delta^{(n)})}
                                {\sinh(\eta(C^{(n)}+2)-u+\delta^{(n)})}
                           \right)\\
\nonumber
&\times & \Bigg\{\prod_{i=1}^{M}\frac{\sinh(u-\lambda_{i}+2\eta)}
                                 {\sinh(u-\lambda_{i})}\,
             \prod_{\alpha=1}^{m}\frac{\sinh(u-\mu_{\alpha}-2\eta)}
                                 {\sinh(u-\mu_{\alpha})}\\
\nonumber
&       & +\qquad\qquad\qquad\qquad\quad\:
          \prod_{\alpha=1}^{m}\frac{\sinh(u-\mu_{\alpha}+2\eta)}
                              {\sinh(u-\mu_{\alpha})}\Bigg\}\\
\nonumber
& - & (-1)^{N}\prod_{i=1}^{M}\frac{\sinh(u-\lambda_{i}+2\eta)}
                                  {\sinh(u-\lambda_{i})}
\end{eqnarray}
The BA-equations (analyticity conditions) are
\begin{eqnarray}
\label{V3-solb}
      \prod_{i=1}^{M}\frac{\sinh(\mu_{\alpha}-\lambda_{i}+2\eta)}
                          {\sinh(\mu_{\alpha}-\lambda_{i})}
& = & \prod_{\stackrel{\s \beta=1}{\s \beta\neq\alpha}}^{m}
                     \frac{\sinh(\mu_{\alpha}-\mu_{\beta}+2\eta)}
                          {\sinh(\mu_{\alpha}-\mu_{\beta}-2\eta)}
\end{eqnarray}
for $\alpha =1,\ldots,m$ and
\begin{eqnarray}
\nonumber
&   & \prod_{n=1}^{N}\frac{\sinh(\lambda_{k}-\delta^{(n)}-\eta(C^{(n)}+2))}
                          {\sinh(\lambda_{k}-\delta^{(n)}+\eta C^{(n)})}\\
\label{V3-solc}
& = & \prod_{\alpha =1}^{m}\frac{\sinh(\mu_{\alpha}-\lambda_{k}+2\eta)}
                          {\sinh(\mu_{\alpha}-\lambda_{k})}
\end{eqnarray}
\end{mathletters}
for $k=1,\ldots,M$. The situation is different in the case of 
$\tau^{V_{4}}(N|u)$, because the innocent looking change of auxiliary space 
requires the use of an at first sight completely different algebra. In the 
next section a systematic approach to this problem will be developed, which 
makes extensive use of the presented solution.
\section{Diagonalization of $\hat{\tau}^{V^{\prime}_{4}}(N|u)$}
\label{tau-diag}
In order to understand the difficulties in diagonalizing 
the homogeneous version of $\tau^{V^{\prime}_{4}}(N|u)$ defined in section 
\ref{models}, it is convenient to follow the standard procedure from the 
previous section as far as possible. So 
$V^{(N)}_{4}\times\cdots\times V^{(1)}_{4}$ will be chosen as quantum space, 
while $V^{\prime}_{4}$, characterized by primed parameters (\ref{V4para}) 
will serve as auxiliary space.\\
The sign change (\ref{rescal}) will be applied and the local vacuum 
will be chosen as lowest weight state in $V_{4}$ (\ref{l-vac}). Omitting
the local index $(n)$, due to homogeneity, this leads to
\begin{eqnarray}
\nonumber
&   & \hat{L}^{V^{\prime}_{4}}(n|u)\,\Omega^{(n)}\\
\label{l-vac-act2}
& = & \left(\begin{array}{cccc}
      \omega_{1}(u) & 0 & 0 & 0\\
      \ast & \omega_{2}(u) & \ast & \ast\\
      \ast & 0 & \omega_{3}(u) & 0\\
      \ast & 0 & 0 & \omega_{4}(u)
      \end{array}\right)\,\Omega^{(n)}
\end{eqnarray}
with the new (local) vacuum eigenvalues 
\begin{eqnarray}
\nonumber
\omega_{1}(u) & = & \frac{\sinh(\eta(C-C^{\prime})+u)}
                         {\sinh(\eta(C+C^{\prime})-u)}
                    \frac{\sinh(\eta(C-C^{\prime}+2)+u)}
                         {\sinh(\eta(C+C^{\prime}+2)+u)}\\
\label{V4-vac-ev}
\omega_{2}(u) & = & \frac{\sinh(\eta(C+C^{\prime}+2)-u)}
                         {\sinh(\eta(C+C^{\prime}+2)+u)}\\
\nonumber
\omega_{3}(u) & = & \frac{\sinh(\eta(C^{\prime}-C)-u)}
                         {\sinh(\eta(C+C^{\prime}+2)+u)}\\
\nonumber 
\omega_{4}(u) & = & \omega_{3}(u)
\end{eqnarray}
There are five non-vanishing entries compared to two in (\ref{l-vac-act1}).
This will be the same for the other three possible local vacua.
Using (\ref{g-vac}), (\ref{T-matrix}) leads to
\begin{eqnarray}
\label{g-vac-act2}
&   & \hat{T}^{V^{\prime}_{4}}(N|u)\,\left|0\right>_{N}\\
\nonumber
& = & \left(\begin{array}{cccc}
      [\omega_{1}(u)]^{N} & 0 & 0 & 0\\
      \ast & [\omega_{2}(u)]^{N} & \ast & \ast\\
      \ast & 0 & [\omega_{3}(u)]^{N} & 0\\
      \ast & 0 & 0 & [\omega_{3}(u)]^{N}
      \end{array}\right)\,\left|0\right>_{N}
\end{eqnarray}
>From the integrability condition (\ref{Integr})
\[ \left[\hat{\tau}^{V^{\prime}_{4}}(N|u),\hat{\tau}^{V_{3}}(N|v)\right]
   \:=\:0 \]
it is clear, that $\hat{\tau}^{V_{3}}(N|v)$ and 
$\hat{\tau}^{V^{\prime}_{4}}(N|u)$ share the same eigenvectors. The
eigenvalues (\ref{V3-ev1}) are in general degenerate. The lowest weight 
property of the (global) vacuum (\ref{g-vac}), which is inherited by the 
BA-vectors (\ref{NABA-es}) via standard arguments \cite{QISM}, guarantees 
uniqueness of these special vectors. Note that the same argument would
hold also for a highest weight state as (global) vacuum, but not for any 
other choice. From this and (\ref{g-vac-act2}), following Baxter \cite{Baxt}, 
it can be concluded immediately, that all eigenvalues of 
$\hat{\tau}^{V^{\prime}_{4}}(N|u)$ can be represented in the form
\begin{eqnarray}
\nonumber
&   & \Lambda^{V^{\prime}_{4}}_{N}(u;\lambda_{1},\ldots,\lambda_{M}|
                                     \mu_{1},\ldots,\mu_{m})\\
\label{ev-ansatz}
& = & [\omega_{1}(u)]^{N}F(u)\,+\,[\omega_{2}(u)]^{N}G(u)\\
\nonumber
& - & [\omega_{3}(u)]^{N}\left\{ H(u)+J(u)\right\}
\end{eqnarray}
where $F(u),G(u),H(u)$ and $J(u)$ are meromorphic functions in $u$, whose
residua cancel, if the analyticity conditions (\ref{BA-equ1}) hold.\\
In order to determine these unknown functions, the FCR (\ref{FCR}) with
with $V=V_{3}$ and $V^{\prime}=V^{\prime}_{4}$, namely
\begin{eqnarray}
\nonumber
&   & R^{V_{3}V^{\prime}_{4}}_{12}(u,v)\hat{T}^{V_{3}}_{1}(N|u)
                                       \hat{T}^{V^{\prime}_{4}}_{2}(N|v)\\
\label{FCR4}
& = & \hat{T}^{V^{\prime}_{4}}_{2}(N|v)
      \hat{T}^{V_{3}}_{1}(N|u)R^{V_{3}V^{\prime}_{4}}_{12}(u,v)
\end{eqnarray}
with $R^{V_{3}V^{\prime}_{4}}(u)$ from (\ref{RV3V4}) will be chosen. The
reasons are
\begin{enumerate}
\item $R^{V_{3}V^{\prime}_{4}}$ is a $12\times 12$-matrix while 
      $R^{V_{4}V^{\prime}_{4}}$ is a $16\times 16$-matrix. 
      The choice $V=V_{4}$ would greatly increase the number of
      equations.
\item In contrast to (\ref{g-vac-act1}) equation (\ref{g-vac-act2}) does not 
      offer a natural choice of \emph{creation-operators}, so the invaluable a 
      priori knowledge of unique eigenvectors (\ref{NABA-es}) with 
      BA-parameters obeying (\ref{BA-equ1}) would be lost within the 
      alternative choice.
\end{enumerate}
The $R$-matrices $R^{V_{3}V^{\prime}_{4}}(u)$ (\ref{RV3V4}) and 
$R^{V_{4}V^{\prime}_{4}}(u)$ (\ref{RV4V4}) do not contain 
$R^{V_{2}V_{2}}(u)$ (\ref{R6V}) as a proper submatrix.\\
In particular \emph{unwanted terms} turn out to be much more complicated. 
However it is possible to omit their calculation. As will be shown, the 
knowledge of unique eigenvectors (\ref{NABA-es}) with (\ref{BA-equ1}) as well 
as some details of the calculation given in section \ref{ABA} are
sufficient to determine the unknown functions in (\ref{ev-ansatz}) 
unambiguously.\\
For brevity (\ref{C-op}) will be used as well as
\[ \hat{T}^{V_{3}}_{ij}(u)\:=\:[\hat{T}^{V_{3}}(N|u)]^{V_{3}}_{ij}
   \quad\mbox{,}\quad\hat{T}^{V^{\prime}_{4}}_{ij}(u)
   \:=\:[\hat{T}^{V^{\prime}_{4}}(N|u)]^{V^{\prime}_{4}}_{ij} \]
First it is convenient to list all components from (\ref{FCR4}), containing an 
operator $\hat{C}_{i}(u)$ (\ref{C-op}) multiplied with a diagonal element of
$\hat{T}^{V^{\prime}_{4}}_{jj}(v)$ from the right. From (\ref{RV3V4})
and (\ref{V3V4-coeff}) with primed parameters (\ref{V4para}) and
(\ref{FCR4}) follows:
\begin{mathletters}
\label{FCR5}
\begin{eqnarray}
\nonumber
\zeta_{2}(u-v)\hat{T}^{V_{3}}_{11}(u)\hat{T}^{V^{\prime}_{4}}_{41}(v)& &\\
\label{FCR5a}
-\,\zeta_{2}(u-v)q\hat{T}^{V_{3}}_{21}(u)\hat{T}^{V^{\prime}_{4}}_{31}(v) & &\\
\nonumber
+\,\beta_{0}(u-v)\hat{C}_{1}(u)\hat{T}^{V^{\prime}_{4}}_{11}(v)
& = & \rho(u-v)\hat{T}^{V^{\prime}_{4}}_{11}(v)\hat{C}_{1}(u)\,,
\end{eqnarray}
\begin{eqnarray}
\label{FCR5b}
-\hat{C}_{1}(u)\hat{T}^{V^{\prime}_{4}}_{22}(v)
& = & \alpha_{0}(u-v)\hat{T}^{V^{\prime}_{4}}_{22}(v)\hat{C}_{1}(u)\\
\nonumber
& - & \varepsilon_{2}(u-v)\hat{T}^{V^{\prime}_{4}}_{23}(v)
      \hat{T}^{V_{3}}_{33}(u)\,,
\end{eqnarray}
\begin{eqnarray}
\nonumber
\varepsilon_{2}(u-v)\hat{T}^{V_{3}}_{11}(u)
\hat{T}^{V^{\prime}_{4}}_{23}(v) & & \\
\label{FCR5c}
+\,\gamma_{0}(u-v)\hat{C}_{1}(u)\hat{T}^{V^{\prime}_{4}}_{33}(v)
& = & \rho(u-v)\hat{T}^{V^{\prime}_{4}}_{33}(v)\hat{C}_{1}(u)\,,
\end{eqnarray}
\begin{eqnarray}
\nonumber
\varepsilon_{2}(u-v)\hat{T}^{V_{3}}_{21}(u)
\hat{T}^{V^{\prime}_{4}}_{24}(v)& &\\
\label{FCR5d}
+\,\gamma_{0}(u-v)\hat{C}_{1}(u)\hat{T}^{V^{\prime}_{4}}_{44}(v)
& = & \alpha_{0}(u-v)\hat{T}^{V^{\prime}_{4}}_{44}(v)\hat{C}_{1}(u)\,\\
\nonumber
& + & \delta_{2}(u-v)\hat{T}^{V^{\prime}_{4}}_{43}(v)\hat{C}_{2}(u)\\
\nonumber
& - & \zeta_{2}(u-v)\hat{T}^{V^{\prime}_{4}}_{41}(v)
      \hat{T}^{V_{3}}_{33}(u)
\end{eqnarray}
\begin{eqnarray}
\nonumber
\zeta_{2}(u-v)\hat{T}^{V_{3}}_{12}(u)\hat{T}^{V^{\prime}_{4}}_{41}(v) & & \\
\label{FCR5e}
-\,\zeta_{2}(u-v)q\hat{T}^{V_{3}}_{22}(u)
                  \hat{T}^{V^{\prime}_{4}}_{31}(v) & & \\
\nonumber
+\,\beta_{0}(u-v)\hat{C}_{2}(u)\hat{T}^{V^{\prime}_{4}}_{12}(v)
& = & \rho(u-v)\hat{T}^{V^{\prime}_{4}}_{11}(v)\hat{C}_{2}(u)
\end{eqnarray}
\begin{eqnarray}
\label{FCR5f}
      -\hat{C}_{2}(u)\hat{T}^{V^{\prime}_{4}}_{22}(v)
& = & \alpha_{0}(u-v)\hat{T}^{V^{\prime}_{4}}_{22}\hat{C}_{2}(u)\\
\nonumber
& - & \varepsilon_{2}(u-v)\hat{T}^{V^{\prime}_{4}}_{24}\hat{T}^{V_{3}}_{33}(v)
\end{eqnarray}
\begin{eqnarray}
\nonumber
\varepsilon_{2}(u-v)\hat{T}^{V_{3}}_{12}(u)
                    \hat{T}^{V^{\prime}_{4}}_{23}(v) &  & \\
\label{FCR5g}
+\,\gamma_{0}(u-v)\hat{C}_{2}(u)\hat{T}^{V^{\prime}_{4}}_{33}(v)
& = & \delta_{1}(u-v)\hat{T}^{V^{\prime}_{4}}_{34}(v)\hat{C}_{1}(u)\\
\nonumber
& + & \alpha_{0}(u-v)\hat{T}^{V^{\prime}_{4}}_{33}(v)\hat{C}_{2}(u)\\
\nonumber
& + & \zeta_{2}(u-v)q\hat{T}^{V^{\prime}_{4}}_{31}(v)\hat{T}^{V_{3}}_{33}(u)
\end{eqnarray}
\begin{eqnarray}
\nonumber
\varepsilon_{2}(u-v)\hat{T}^{V_{3}}_{22}(u)
                    \hat{T}^{V^{\prime}_{4}}_{24}(u) &  & \\
\label{FCR5h}
+\,\gamma_{0}(u-v)\hat{C}_{2}(u)\hat{T}^{V^{\prime}_{4}}_{44}(v)
& = & \rho(u-v)\hat{T}^{V^{\prime}_{4}}_{44}(v)\hat{C}_{2}(u)
\end{eqnarray}
\end{mathletters}
The idea is to keep only contributions \emph{leading to wanted terms},
when the eigenvector (\ref{NABA-es}) is applied to 
$\hat{\tau}^{V^{\prime}_{4}}(N|u)$ and neglect all others. 
The set (\ref{FCR5}) is not complete. For instance in (\ref{FCR5a}) a term 
$\propto\hat{T}^{V_{3}}_{11}(u)\hat{T}^{V^{\prime}_{4}}_{41}$
and another $\propto\hat{T}^{V_{3}}_{21}(u)\hat{T}^{V^{\prime}_{4}}_{31}$ 
occur. Both will act non-trivially on $\left|0\right>_{N}$ from 
(\ref{g-vac-act2}). However in the set (\ref{FCR4}) the relations 
\begin{eqnarray*}
\alpha_{0}(u-v)\hat{T}^{V_{3}}_{11}(u)\hat{T}^{V^{\prime}_{4}}_{41}(v) & & \\
+\,\delta_{1}(u-v)\hat{T}^{V_{3}}_{21}(u)\hat{T}^{V^{\prime}_{4}}_{31}(v) & &\\
+\,\zeta_{1}(u-v)\hat{C}_{1}(u)\hat{T}^{V^{\prime}_{4}}_{11}(v)
& = & \hat{T}^{V^{\prime}_{4}}_{41}(v)\hat{T}^{V_{3}}_{11}(u)
\end{eqnarray*}
and
\begin{eqnarray*}
\delta_{2}(u-v)\hat{T}^{V_{3}}_{11}(u)\hat{T}^{V^{\prime}_{4}}_{41}(v) &  & \\
+\,\alpha_{0}(u-v)\hat{T}^{V_{3}}_{21}(u)\hat{T}^{V^{\prime}_{4}}_{31}(v)& &\\
  -\, \zeta_{1}(u-v)q^{-1}\hat{C}_{1}(u)\hat{T}^{V^{\prime}_{4}}_{11}(v)
& = & \rho(u-v)\hat{T}^{V^{\prime}_{4}}_{31}(v)\hat{T}^{V_{3}}_{21}(u)
\end{eqnarray*}
can be found and used to eliminate these terms leading to
\begin{eqnarray*}
&   & \left(\beta_{0} -
            \frac{\zeta_{1}\zeta_{2}
                  [2\alpha_{0}+q^{-1}\delta_{1}+q\delta_{2}]}
                 {\alpha^{2}_{0}-\delta_{1}\delta_{2}}\right)(u-v)\,
      \hat{C}_{1}(u)\hat{T}^{V^{\prime}_{4}}_{11}(v)\\
& = & \rho(u-v)\,\hat{T}^{V^{\prime}_{4}}_{11}(v)\hat{C}_{1}(u)\\
& - & \left(\frac{\rho\zeta_{2}}{\alpha_{0}}
            \left[1+\frac{\delta_{2}\left[\alpha_{0} q+\delta_{1}\right]}
                         {\alpha^{2}_{0}-\delta_{1}\delta_{2}}\right]
      \right)(u-v)\,\hat{T}^{V^{\prime}_{4}}_{41}(v)\hat{T}^{V_{3}}_{11}(u)\\
& + & \left(\frac{\rho\zeta_{2}[\alpha_{0}q+\delta_{1}]}
                 {\alpha^{2}_{0}-\delta_{1}\delta_{2}}\right)(u-v)\,
      \hat{T}^{V^{\prime}_{4}}_{31}\hat{T}^{V_{3}}_{21}(u)
\end{eqnarray*}
where the dependence on difference variables has been denoted symbolically
for brevity. The last two terms on the right hand side will not lead to a 
contribution proportional to any BA-eigenvector (\ref{NABA-es}). It has been 
checked -- and this is crucial, that these terms are not related to a proper 
combination of $\hat{C}_{i}$-operators by unused relations from the set 
(\ref{FCR4}). In conclusion they can be identified as 
\emph{leading to unwanted terms}.\\
In the same way two other relations from (\ref{FCR4}) may be used to 
eliminate from (\ref{FCR5e}) terms $\propto\hat{T}^{V_{3}}_{12}(u)
\hat{T}^{V^{\prime}_{4}}_{41}(v)$ and $\propto\hat{T}^{V_{3}}_{22}(u)
\hat{T}^{V^{\prime}_{4}}_{31}(v)$, which after 
omitting contributions \emph{leading to unwanted terms} yield the 
same result with $\hat{C}_{1}(u)$ replaced by $\hat{C}_{2}(u)$, i.e.:
\begin{mathletters}
\label{FCR6}
\begin{eqnarray}
\nonumber
      \hat{T}^{V^{\prime}_{4}}_{11}(u)\hat{C}_{i}(v) 
& = & \left(\frac{\beta_{0}}{\rho} -
            \frac{\zeta_{1}\zeta_{2}
                  [2\alpha_{0}+q^{-1}\delta_{1}+q\delta_{2}]}
                 {\rho[\alpha^{2}_{0}-\delta_{1}\delta_{2}]}\right)(v-u)\\
\label{FCR6a}
&\times & \hat{C}_{i}(v)\hat{T}^{V^{\prime}_{4}}_{11}(u)\quad\pm\:\ldots
\end{eqnarray}
for $i=1,2.$\\
In (\ref{FCR5b}) and (\ref{FCR5f}) terms $\propto\hat{T}^{V^{\prime}_{4}}(v)
\hat{T}^{V_{3}}_{33}(u)$ and $\propto\hat{T}^{V^{\prime}_{4}}_{24}
\hat{T}^{V_{3}}_{33}(v)$ can be identified as \emph{leading to unwanted terms}
in the sense explained above and therefore be neglected:
\begin{eqnarray}
\label{FCR6b}
      \hat{T}^{V^{\prime}_{4}}_{22}(u)\hat{C}_{i}(v)
& = & \frac{-1}{\alpha_{0}(v-u)}\,
      \hat{C}_{i}(v)\hat{T}^{V^{\prime}_{4}}_{22}(u)
      \quad\pm\:\ldots
\end{eqnarray}
for $i=1,2$. The other relations from (\ref{FCR4}) can be treated similarly,
leading to
\begin{eqnarray}
\nonumber
      \hat{T}^{V^{\prime}_{4}}_{33}(u)\hat{C}_{1}(v)
& = & \left(\frac{\alpha_{0}\gamma_{0}-\varepsilon_{1}\varepsilon_{2}}
                 {\alpha_{0}\rho}\right)(v-u)\,
      \hat{C}_{1}(v)\hat{T}^{V^{\prime}_{4}}_{33}(u)\\
\label{FCR6c}
&\pm& \ldots\\
\nonumber
      \hat{T}^{V^{\prime}_{4}}_{44}(u)\hat{C}_{2}(v)
& = & \left(\frac{\alpha_{0}\gamma_{0}-\varepsilon_{1}\varepsilon_{2}}
                 {\alpha_{0}\rho}\right)(v-u)\,
      \hat{C}_{2}(v)\hat{T}^{V^{\prime}_{4}}_{44}(u)\\
\label{FCR6d}
&\pm& \ldots\\
\nonumber
      \hat{T}^{V^{\prime}_{4}}_{44}(u)\hat{C}_{1}(v)
& = & \left(\frac{\alpha_{0}\gamma_{0}-\varepsilon_{1}\varepsilon_{2}}
                 {\alpha^{2}_{0}-\delta_{1}\delta_{2}}\right)(v-u)\,
      \hat{C}_{1}(v)\hat{T}^{V^{\prime}_{4}}_{44}(u)\\
\nonumber
& - & \left(\frac{\delta_{2}}{\alpha_{0}}
            \frac{\alpha_{0}\gamma_{0}-\varepsilon_{1}\varepsilon_{2}}
                 {\alpha^{2}_{0}-\delta_{1}\delta_{2}}\right)(v-u)\,
      \hat{C}_{2}(v)\hat{T}^{V^{\prime}_{4}}_{43}(u)\\
\label{FCR6e}
&\pm& \ldots\\
\nonumber
      \hat{T}^{V^{\prime}_{4}}_{33}(u)\hat{C}_{2}(v)
& = & \left(\frac{\alpha_{0}\gamma_{0}-\varepsilon_{1}\varepsilon_{2}}
                 {\alpha^{2}_{0}-\delta_{1}\delta_{2}}\right)(v-u)\,
      \hat{C}_{2}(v)\hat{T}^{V^{\prime}_{4}}_{33}(u)\\
\nonumber
& - & \left(\frac{\delta_{1}}{\alpha_{0}}
            \frac{\alpha_{0}\gamma_{0}-\varepsilon_{1}\varepsilon_{2}}
                 {\alpha^{2}_{0}-\delta_{1}\delta_{2}}\right)(v-u)\,
      \hat{C}_{1}(v)\hat{T}^{V^{\prime}_{4}}_{43}(u)\\
\label{FCR6f}
&\pm& \ldots
\end{eqnarray}
\end{mathletters}
Some details of the calculations are given in appendix \ref{AppC}. 
They are tedious, but straightforward: It is trivial to identify
terms proportional to a simple $M=1$ eigenvector, (\ref{NABA-es}), if it is 
applied. The remaining terms are divided into those, which possibly lead to a
contribution proportional to an eigenvector via the algebra (\ref{FCR5}), and 
others which cannot be transformed this way. The former terms have been 
eliminated by using convenient relations from (\ref{FCR5}) and evaluated 
again, till this procedure terminated, leaving only terms of the latter type, 
i.e. \emph{unwanted terms}, which have been neglected systematically in 
(\ref{FCR6}).\\
Equations (\ref{FCR6e}) and (\ref{FCR6f}) contain non-trivial terms 
$\propto\hat{C}_{2}(v)\hat{T}^{V^{\prime}_{4}}_{43}(u)$ and
$\propto\hat{C}_{1}(v)\hat{T}^{V^{\prime}_{4}}_{43}(u)$. Next
it is natural to add to (\ref{FCR5}) the relations involving terms
$\propto\hat{T}^{V^{\prime}_{4}}_{34}(u)\hat{C}_{i}(v)$ and 
$\propto\hat{T}^{V^{\prime}_{4}}_{43}(u)\hat{C}_{i}(v)$ with $i=1,2$, i.e.
\begin{mathletters}
\label{FCR7}
\begin{eqnarray}
\nonumber
      \varepsilon_{2}(u-v)\hat{T}^{V_{3}}_{11}(u)
      \hat{T}^{V^{\prime}_{4}}_{24}(v) & & \\
\label{FCR7a}
  +\, \gamma_{0}(u-v)\hat{C}_{1}(u)\hat{T}^{V^{\prime}_{4}}_{34}(v)
& = & \alpha_{0}(u-v)\hat{T}^{V^{\prime}_{4}}_{34}(v)\hat{C}_{1}(u)\\
\nonumber
& + & \delta_{2}(u-v)\hat{T}^{V^{\prime}_{4}}_{33}(v)\hat{C}_{2}(u)\\
\nonumber
& + & \zeta_{2}(u-v)\hat{T}^{V^{\prime}_{4}}_{31}(v)\hat{T}^{V_{3}}_{33}(u)
\end{eqnarray}
\begin{eqnarray}
\nonumber
      \varepsilon_{2}(u-v)\hat{T}^{V_{3}}_{21}(u)
      \hat{T}^{V^{\prime}_{4}}_{23}(v) & & \\
\label{FCR7b}
  +\, \gamma_{0}(u-v)\hat{C}_{1}(u)\hat{T}^{V^{\prime}_{4}}_{43}(v)
& = & \rho(u-v)\hat{T}^{V^{\prime}_{4}}_{43}(v)\hat{C}_{1}(u)
\end{eqnarray}
\begin{eqnarray}
\nonumber
      \varepsilon_{2}(u-v)\hat{T}^{V_{3}}_{22}(u)
      \hat{T}^{V^{\prime}_{4}}_{24}(v) & & \\
\label{FCR7c}
  +\, \gamma_{0}(u-v)\hat{C}_{2}(u)\hat{T}^{V^{\prime}_{4}}_{34}(v)
& = & \rho(u-v)\hat{T}^{V^{\prime}_{4}}_{34}(v)\hat{C}_{2}(u)
\end{eqnarray}
\begin{eqnarray}
\nonumber
      \varepsilon_{2}(u-v)\hat{T}^{V_{3}}_{22}(u)
      \hat{T}^{V^{\prime}_{4}}_{23}(v) & & \\
\label{FCR7d}
  +\, \gamma_{0}(u-v)\hat{C}_{2}(u)\hat{T}^{V^{\prime}_{4}}_{43}(v)
& = & \delta_{1}(u-v)\hat{T}^{V^{\prime}_{4}}_{44}(v)\hat{C}_{1}(u)\\
\nonumber
& + & \alpha_{0}(u-v)\hat{T}^{V^{\prime}_{4}}_{43}(v)\hat{C}_{2}(u)\\
\nonumber
& + & \zeta_{2}(u-v)q\hat{T}^{V^{\prime}_{4}}_{41}(v)\hat{T}^{V_{3}}_{33}(u)
\end{eqnarray}
\end{mathletters}
Proceeding as above, leads to 
\begin{mathletters}
\label{FCR8}
\begin{eqnarray}
\nonumber
      \hat{T}^{V^{\prime}_{4}}_{34}(u)\hat{C}_{1}(v)
& = & \left(\frac{\alpha_{0}\gamma_{0}-\varepsilon_{1}\varepsilon_{2}}
                 {\alpha^{2}_{0}-\delta_{1}\delta_{2}}\right)(v-u)\,
      \hat{C}_{1}(v)\hat{T}^{V^{\prime}_{4}}_{34}(u)\\
\nonumber
& - & \left(\frac{\delta_{2}}{\alpha_{0}}
            \frac{\alpha_{0}\gamma_{0}-\varepsilon_{1}\varepsilon_{2}}
                 {\alpha^{2}_{0}-\delta_{1}\delta_{2}}\right)(v-u)\,
      \hat{C}_{2}(v)\hat{T}^{V^{\prime}_{4}}_{33}(u)\\
\label{FCR8a}
&\pm& \ldots\\
\nonumber
      \hat{T}^{V^{\prime}_{4}}_{43}(u)\hat{C}_{2}(v)
& = & \left(\frac{\alpha_{0}\gamma_{0}-\varepsilon_{1}\varepsilon_{2}}
                 {\alpha^{2}_{0}-\delta_{1}\delta_{2}}\right)(v-u)\,
      \hat{C}_{2}(v)\hat{T}^{V^{\prime}_{4}}_{43}(u)\\
\nonumber
& - & \left(\frac{\delta_{1}}{\alpha_{0}}
            \frac{\alpha_{0}\gamma_{0}-\varepsilon_{1}\varepsilon_{2}}
                 {\alpha^{2}_{0}-\delta_{1}\delta_{2}}\right)(v-u)\,
      \hat{C}_{1}(v)\hat{T}^{V^{\prime}_{4}}_{44}(u)\\
\label{FCR8b}
&\pm& \ldots\\
\nonumber
      \hat{T}^{V^{\prime}_{4}}_{43}(u)\hat{C}_{1}(v)
& = & \left(\frac{\alpha_{0}\gamma_{0}-\varepsilon_{1}\varepsilon_{2}}
                 {\alpha_{0}\rho}\right)(v-u)\,
      \hat{C}_{1}(v)\hat{T}^{V^{\prime}_{4}}_{43}(u)\\
\label{FCR8c}
&\pm& \ldots\\
\nonumber
      \hat{T}^{V^{\prime}_{4}}_{34}(u)\hat{C}_{2}(v)
& = & \left(\frac{\alpha_{0}\gamma_{0}-\varepsilon_{1}\varepsilon_{2}}
                 {\alpha_{0}\rho}\right)(v-u)\,
      \hat{C}_{2}(v)\hat{T}^{V^{\prime}_{4}}_{34}(u)\\
\label{FCR8d}
&\pm& \ldots
\end{eqnarray}
\end{mathletters}
This idea is strongly supported by a comparison of (\ref{FCR6}) and 
(\ref{FCR8}) with (\ref{FCR2}), used in the algebraic diagonalization 
of $\hat{\tau}^{V_{3}}(N|u)$, suggesting, that the submatrix 
$\{\hat{T}^{V^{\prime}_{4}}_{ij}\}$ with $i,j=3,4$ will play the same 
r\^ole as the submatrix $\{\hat{T}^{V_{3}}_{ij}\}$ with $i,j=1,2$ in the 
previous section. Indeed using the definitions (\ref{V3V4-coeff}) with
primed parameters (\ref{V4para}), (\ref{FCR6a}) and (\ref{FCR6b}) can be 
written
\begin{mathletters}
\label{FCR9}
\begin{eqnarray}
\nonumber
      \hat{T}^{V^{\prime}_{4}}_{11}(u)\hat{C}_{i}(v) 
& = & \frac{\sinh(u-v+\eta(C^{\prime}+2))}{\sinh(u-v-\eta(C^{\prime}-2))}
      \hat{C}_{i}(v)\hat{T}^{V^{\prime}_{4}}_{11}(u)\\
\label{FCR9a}
&\pm& \ldots
\end{eqnarray}
\begin{eqnarray}
\nonumber
      \hat{T}^{V^{\prime}_{4}}_{22}(u)\hat{C}_{i}(v) 
& = & \frac{\sinh(u-v+\eta(C^{\prime}+2))}{\sinh(u-v-\eta C^{\prime})}
      \hat{C}_{i}(v)\hat{T}^{V^{\prime}_{4}}_{22}(u)\\
\label{FCR9b}
&\pm& \ldots
\end{eqnarray}
for $i=1,2.$, while the remaining equations from (\ref{FCR6}) and 
(\ref{FCR8}) may be noted as
\begin{eqnarray}
\nonumber
      \hat{t}_{ij}(u)\hat{C}_{k}(v) 
& = & \frac{\sinh(u-v+\eta(C^{\prime}+2))}{\sinh(u-v-\eta C^{\prime})}\\
\nonumber
&\times & \sum_{l,m=1}^{2}r_{lm,jk}(u-v-\eta C^{\prime})\,
          \hat{C}_{m}(v)\hat{t}_{il}(u)\\
\label{FCR9c}
&\pm& \ldots
\end{eqnarray}
for $i,j,k=1,2$, where the elements $r_{ik,jl}(u)$ of the $R$-matrix 
(\ref{R6V}) and the convenient definition
\begin{eqnarray}
\label{FCR9d}
\left(\begin{array}{cc}
      \hat{t}_{11}(u) & \hat{t}_{12}(u)\\
      \hat{t}_{21}(u) & \hat{t}_{22}(u)
      \end{array}\right) & := & 
\left(\begin{array}{cc}
      \hat{T}^{V^{\prime}_{4}}_{33}(u) & \hat{T}^{V^{\prime}_{4}}_{43}(u)\\
      \hat{T}^{V^{\prime}_{4}}_{34}(u) & \hat{T}^{V^{\prime}_{4}}_{44}(u)
      \end{array}\right)
\end{eqnarray}
\end{mathletters}
have been used. The similarity of (\ref{FCR9}) to (\ref{FCR2}) is striking and 
allows to calculate the eigenvalues of $\hat{\tau}^{V^{\prime}_{4}}(N|u)$
easily.\\
Applying the (right) eigenvector (\ref{NABA-es}) to 
$\hat{T}^{V^{\prime}_{4}}_{11}(u)$ and
$\hat{T}^{V^{\prime}_{4}}_{22}(u)$ using (\ref{FCR9}) and 
(\ref{g-vac-act2}) yields
\begin{eqnarray*}
&   & \hat{T}^{V^{\prime}_{4}}_{11}(u)
      |\lambda_{1},\ldots,\lambda_{M}|F>\\
& = & [\omega_{1}(u)]^{N}\prod_{i=1}^{M}
      \frac{\sinh(u-\lambda_{i}+\eta(C^{\prime}+2))}
           {\sinh(u-\lambda_{i}-\eta(C^{\prime}-2))}\\
&\times& |\lambda_{1},\ldots,\lambda_{M}|F>\quad\pm\:\ldots
\end{eqnarray*}
and
\begin{eqnarray*}
&   & \hat{T}^{V^{\prime}_{4}}_{22}(u)
      |\lambda_{1},\ldots,\lambda_{M}|F>\\
& = & [\omega_{2}(u)]^{N}\prod_{i=1}^{M}
      \frac{\sinh(u-\lambda_{i}+\eta(C^{\prime}+2))}
           {\sinh(u-\lambda_{i}-\eta C^{\prime})}\\
&\times& |\lambda_{1},\ldots,\lambda_{M}|F>\quad\pm\:\ldots
\end{eqnarray*}
where \emph{unwanted terms} have been omitted. Applying it to 
$[\hat{T}^{V^{\prime}_{4}}_{33}(u)+\hat{T}^{V^{\prime}_{4}}_{44}(u)]$
yields
\begin{eqnarray*}
&   & \left[\hat{T}^{V^{\prime}_{4}}_{33}(u)+
            \hat{T}^{V^{\prime}_{4}}_{44}(u)\right]
      |\lambda_{1},\ldots,\lambda_{M}|F>\\
& = & [\omega_{3}(u)]^{N}\prod_{i=1}^{M}
      \frac{\sinh(u-\lambda_{i}+\eta(C^{\prime}+2))}
           {\sinh(u-\lambda_{i}-\eta C^{\prime})}\\
&\times& \left[\hat{\tau}^{V_{2}}(M|u-\eta C^{\prime})
         \right]^{b_{1},\ldots,b_{M}}_{a_{1},\ldots,a_{M}}
         F^{a_{1},\ldots,a_{M}}\\
&\times& \hat{C}_{b_{1}}(\lambda_{1})\cdots\hat{C}_{b_{M}}(\lambda_{M})
         \left|0\right>_{N}\quad\pm\:\ldots
\end{eqnarray*}
where $\hat{\tau}^{V_{2}}(M|u)$ is defined by (\ref{R6V}) via (\ref{QISM1}) 
with $\delta^{(n)}=\lambda_{n}$ as in section \ref{ABA}. But $F$ is a 
(right) eigenvector to $\hat{\tau}^{V_{2}}(M|u)$ corresponding 
to the eigenvalue from (\ref{6V-ev}). The neglected \emph{unwanted terms} 
vanish per construction if the supertrace (\ref{s-tr}) is performed 
according to (\ref{t-matrix}). Therefore the eigenvalue of 
$\hat{\tau}^{V^{\prime}_{4}}(M|u)$ corresponding to the (right) eigenvector 
(\ref{NABA-es}) is given by
\begin{eqnarray}
\nonumber
&   & \Lambda^{V^{\prime}_{4}}_{N}(u;\lambda_{1},\ldots,\lambda_{M}|
                                     \mu_{1},\ldots,\mu_{m})\\
\label{V4-ev}
& = & [\omega_{1}(u)]^{N}\left(\prod_{i=1}^{M}
      \frac{\sinh(u-\lambda_{i}+\eta(C^{\prime}+2))}
           {\sinh(u-\lambda_{i}-\eta(C^{\prime}-2))}\right)\\
\nonumber
& + & [\omega_{2}(u)]^{N}\left(\prod_{i=1}^{M}
      \frac{\sinh(u-\lambda_{i}+\eta(C^{\prime}+2))}
      {\sinh(u-\lambda_{i}-\eta C^{\prime})}\right)\\
\nonumber
& - & [\omega_{3}(u)]^{N}\left(\prod_{i=1}^{M}
      \frac{\sinh(u-\lambda_{i}+\eta(C^{\prime}+2))}
           {\sinh(u-\lambda_{i}-\eta C^{\prime})}\right)\\
\nonumber
&   & \qquad\quad
      \times\Lambda^{V_{2}}_{M}(u-\eta C^{\prime};\mu_{1},\ldots,\mu_{m})
\end{eqnarray}
with vacuum eigenvalues $\omega_{i}(u)$ ($i=1,2,3$) from (\ref{V4-vac-ev}) and 
$\Lambda^{V_{2}}_{M}(u;\ldots)$ from (\ref{6V-ev}).\\
The BA-parameters $\lambda_{1},\ldots,\lambda_{M}$ and 
$\mu_{1},\ldots,\mu_{m}$ are to be determined by the BA-equations 
(\ref{BA-equ1}). Note, that these are necessary and sufficient conditions 
\cite{Baxt} for analyticity of the eigenvalues (\ref{V4-ev}) 
in $u$. Since up to now no explicit use has been made of these, this is a 
valuable consistency check on the validity of (\ref{V4-ev}).\\
(\ref{V4-ev}) is clearly of the expected form (\ref{ev-ansatz}). It is further
obvious, that the eigenvalues for every \emph{transfer-matrix} 
based on auxiliary space $V^{\prime}_{4}$ can be represented by the same
formula (\ref{V4-vac-ev}), provided the (global) quantum space is a lowest 
weight space. Of course the vacuum eigenvalues have to be replaced by new 
ones, which are obviously restricted by the BA-equations
(\ref{BA-equ1}), as discussed in section \ref{ABA}.\\
For completeness the trivial generalization \cite{Baxt} 
of (\ref{V4-ev}) to the inhomogeneous case with 
$w^{(n)}=C^{(n)}$ in (\ref{L-matrix}) and $\delta^{(n)}\neq 0$ in
(\ref{T-matrix}) shall be given explicitly:
\begin{eqnarray}
\nonumber
&   & \Lambda^{V^{\prime}_{4}}_{N}(u;\lambda_{1},\ldots,\lambda_{M}|
                                     \mu_{1},\ldots,\mu_{m})\\
\nonumber
& = & \Bigg[
      \prod_{n=1}^{N}\frac{\sinh(\eta(C^{(n)}-C^{\prime})+u-\delta^{(n)})}
                          {\sinh(\eta(C^{(n)}+C^{\prime})-u+\delta^{(n)})}\\
\nonumber
&   & \qquad\times\frac{\sinh(\eta(C^{(n)}-C^{\prime}+2)+u-\delta^{(n)})}
                       {\sinh(\eta(C^{(n)}+C^{\prime}+2)+u-\delta^{(n)})}
      \Bigg]\\
\nonumber
&\times & \left(\prod_{i=1}^{M}
                \frac{\sinh(u-\lambda_{i}+\eta(C^{\prime}+2))}
                     {\sinh(u-\lambda_{i}-\eta(C^{\prime}-2))}\right)\\
\label{V4-sol}
& + & \Bigg[\prod_{n=1}^{N}
      \frac{\sinh(\eta(C^{(n)}+C^{\prime}+2)-u+\delta^{(n)})}
           {\sinh(\eta(C^{(n)}+C^{\prime}+2)+u-\delta^{(n)})}\Bigg]\\
\nonumber
&\times& \left(\prod_{i=1}^{M}
         \frac{\sinh(u-\lambda_{i}+\eta(C^{\prime}+2))}
              {\sinh(u-\lambda_{i}-\eta C^{\prime})}\right)\\
\nonumber
& - & \Bigg[\prod_{n=1}^{N}
      \frac{\sinh(\eta(C^{\prime}-C^{(n)})-u+\delta^{(n)})}
           {\sinh(\eta(C^{(n)}+C^{\prime}+2)+u-\delta^{(n)})} \Bigg]\\
\nonumber
&\times& \Bigg\{\left(\prod_{i=1}^{M}
         \frac{\sinh(u-\lambda_{i}+\eta(C^{\prime}+2)}
              {\sinh(u-\lambda_{i}-\eta C^{\prime})}\right)\\
\nonumber
&      & \times\left(\prod_{\alpha=1}^{m}
         \frac{\sinh(u-\mu_{\alpha}-\eta(C^{\prime}+2))}
              {\sinh(u-\mu_{\alpha}-\eta C^{\prime})}\right)\\
\nonumber
&   & +  \left(\prod_{i=1}^{M}
         \frac{\sinh(u-\lambda_{i}+\eta(C^{\prime}+2)}
              {\sinh(u-\lambda_{i}-\eta(C^{\prime}-2))}\right)\\
\nonumber
&      & \times\left(\prod_{\alpha=1}^{m}
         \frac{\sinh(u-\mu_{\alpha}-\eta(C^{\prime}-2))}
              {\sinh(u-\mu_{\alpha}-\eta C^{\prime})}\right)\Bigg\}
\end{eqnarray}
Here the BA-parameters $\lambda_{1},\ldots,\lambda_{M}$ and
$\mu_{1},\ldots,\mu_{m}$ are determined by (\ref{V3-solb})
and (\ref{V3-solc}). (\ref{V4-sol}) describes all eigenvalues. 
As mentioned above, additional eigenvectors to the same eigenvalue 
(\ref{V4-sol}) are obtained by applying shift operators, corresponding to 
the representation of the group-symmetry on the (global) quantum space, to the 
eigenvectors (\ref{NABA-es}). Completeness may be assured by 
the usual arguments \cite{QISM}.
\section{Conclusion}
In the previous section $\hat{\tau}^{V^{\prime}_{4}}(N|u)$ has been 
diagonalized by NABA, combined with analyticity arguments. 
Obviously the method can be applied to any BA-integrable model, defined by 
(\ref{QISM1}), based on an arbitrary, but finite dimensional
representation $V^{\prime}$ of a possibly $q$-deformed Lie 
(super-)algebra as auxiliary space.\\
Let the model based on the direct product of a fundamental representation
$V$ with itself, here defined by $R^{V_{3}V_{3}}(u)$ and
(\ref{QISM1}), be solved by (N)ABA. In order to solve the model
under consideration the following scheme may be applied.
\begin{enumerate}
\item An auxiliary model based on $V$ as auxiliary and the non-standard 
      representation $V^{\prime}$ as quantum space, may be constructed
      by standard methods and its \emph{transfer-matrix}, i.e. 
      $\hat{\tau}^{V_{3}}(N|u)$ from (\ref{RV3V3}) via (\ref{QISM1}), may
      be diagonalized, using a (global) lowest or highest weight state, 
      e.g. $\left|0\right>_{N}$ (\ref{g-vac}), as (pseudo-)vacuum.
\item Vacuum eigenvalues may be calculated trivially, see (\ref{V4-vac-ev}). 
      The \emph{transfer-matrix} of the relevant model and the one of the
      auxiliary model commute (\ref{Integr}) and share all 
      BA-eigenvectors, which dictates the form of the eigenvalue 
      equations (\ref{ev-ansatz}).
\item Mixed FCR (\ref{FCR5}), between \emph{creation-operators} from auxiliary 
      model (\ref{C-op}), should be used as follows:
      \begin{enumerate}
      \item FCRs (\ref{FCR5}) between diagonal elements of 
            $\hat{T}^{V^{\prime}}(N|u)$ and \emph{creation-operators}
            multiplied from the right on these (\ref{FCR6}) should be 
            collected. The remaining terms in these equations are 
            classified as \emph{wanted} (leading to terms proportional
            to the known BA-vectors), \emph{unwanted} (not related to 
            \emph{wanted} ones by FCRs) and others.  
      \item Terms of the last category have to be eliminated by use of other
            convenient FCRs. \emph{Unwanted terms} may be neglected 
            in final equations, i.e. (\ref{FCR6}).
      \item Generically the final equations in step (b) involve some
            off-diagonal elements of $\hat{T}^{V^{\prime}}(N|u)$ (\ref{FCR6}). 
            They have to be complemented by all FCRs containing these
            off-diagonal elements, multiplied from the right with
            \emph{creation-operators} (\ref{FCR7}), to which the same
            procedure has to be applied (\ref{FCR8}). 
      \end{enumerate}
\item The relations obtained in step 3 allow the calculation 
      of the eigenvalue equations (\ref{V4-ev}), if they are
      written down conveniently, i.e. like (\ref{FCR9}).
\end{enumerate}
Step one and two are trivial here. Step three is crucial. An unusually 
large number of FCRs (\ref{FCR5}) has to be used, because the mixed 
$R$-matrix (\ref{RV3V4}), does not contain any smaller $R$-matrix like
(\ref{R6V}) as a proper submatrix, which was true e.g. for (\ref{RV3V3}). 
The approach is systematic and avoids a complicated discussion of 
\emph{unwanted terms}. The author has checked in a number of cases, 
that these indeed vanish in the present application, but analyticity of 
the final result (\ref{V4-ev}) is a very strong and usually sufficient test. 
Step four is simple. Some knowledge of the preceding calculations is a
sufficient guideline.\\
A group theoretical background is not necessary, but helpful. 
Definitely needed is a commuting (auxiliary) model, algebraically solvable 
\cite{QISM}, and a unique identification of joint eigenvectors. The theory 
of quantum groups \cite{Drin,ChPr,Yama} provides both. In addition it is 
implicitly assumed, that the algebra defined by the FCRs is complete, i.e. 
if two operators are identical, this information should be encoded within 
the FCRs. This is guaranteed if $R$ has the intertwining property 
\cite{Drin}.\\
The more complicated problem of handling the full set of commutation 
relations of comparable complexity directly, has been tackled more ore 
less exactly a number of times. The algebraic solution of a statistical 
covering model for the one-dimensional Hubbard model, where no
commuting transfer-matrix is known, by Ramos and Martins \cite{RM1}.
Also a diagonalization of an $Y(sp(2,1))$-symmetrical model by the
same authors should be mentioned \cite{RM2}. To the authors knowledge 
no systematic scheme is known and although the eigenvalues are presumably 
correct, the discussion of unwanted terms is not complete in these works.
\\
It is an interesting, but still unsolved question, if solvability of 
some statistical model by $n$-fold NABA implies the existence of a commuting 
transfer-matrix with minimal, that is $(n+1)$-dimensional, auxiliary space?
\\
In the non-graded $U_{q}(\widehat{gl}^{\prime}(N|{\mathcal C}))$-symmetric 
case, the \emph{quantum-determinant}, introduced by Izergin and Korepin 
\cite{IzKo} and recognized by Drinfel'd \cite{Drin} to complete the center of 
this algebra, provides the possibility to construct functional relations 
\cite{KuSk} for the eigenvalues, extended to an ``analytical Bethe ansatz'' by 
Reshetikhin \cite{Resh}. This is more elegant than the present approach, 
but does not generalize to the graded case, because no one-dimensional 
subspace can be separated from a product of \emph{transfer-matrices}.
\\
The \emph{transfer-matrix} $\hat{\tau}^{V^{\prime}_{4}}(N|u)$ has been used
mainly for pedagogical reasons. Minus signs due to grading, even in the
non-graded version \cite{KuSk} prevent a statistical interpretation.
Nevertheless the Hamiltonian limit in the non-difference type 
spectral parameter (\ref{YB-equ1}), as mentioned above, leads to an 
additional, unusual Hamiltonian, which will be discussed elsewhere \cite{Gru2}.
Note that neither $\hat{\tau}^{V_{3}}(N|u)$ nor $\hat{\tau}^{V_{4}}(N|u)$
are hermitian, except if further restrictions are imposed on (\ref{V4para}).
The diagonalization of $\hat{\tau}^{V_{4}}(N|u)$, especially the
result (\ref{V4-sol}), may serve as starting point for calculations on
the thermodynamics of these models in the non-linear integral equation
approach, pioneered by Kl\"{u}mper \cite{Klue}. For a recent application
of this technique see also \cite{JKS}.\\
The eigenvalue-equation for the \emph{transfer-matrix} of some other 
$U_{q}(\widehat{gl}^{\prime}(2,1|{\mathcal C}))$-symmetric models with 
$V^{\prime}_{4}$ as auxiliary and some lowest weight representation 
as quantum space may be written down by replacing the $\omega_{i}(u)$ 
(\ref{V4-vac-ev}) in (\ref{V4-ev}) by new ones.\\
De Vega and Gonz\'{a}les Ruiz \cite{DeGo} and Foerster and Karowski \cite{FoKa}
generalized the ABA calculations of Schultz \cite{Schu} partially 
to non-periodic, integrable boundary conditions. There should be no principal 
problem to combine their techniques with the method presented here.\\
The perhaps most important open question is concerned with the
applicability of the method to models with infinite dimensional
auxiliary space, which was precautiously excluded here. 
\section{Acknowledgment}
This work has been performed within the research program of the
Sonderforschungsbereich 341 (K\"oln-Aachen-J\"ulich). The author thanks 
J. Zittartz and A. Kl\"umper for continuous support, A. Zvyagin,
G. J\"uttner, Y. Kato, A. Kl\"umper and especially A. Fujii for stimulating 
discussions and encouragement. Special thanks goes to A. Kl\"umper for
carefully reading the manuscript and numerous useful suggestions, 
incorporated in the final version. The author would also like to thank
a referee for pointing out reference \cite{Schu} to him. 
\appendix
\section{Coefficients of R-matrices}
\label{AppA}
The elements of the $R$-matrix (\ref{RV3V4}) are explicitly given by
\begin{mathletters}
\label{V3V4-coeff}
\begin{eqnarray}
\rho(u)            & := & \frac{\sinh(\eta(C+2)+u)}{\sinh(\eta(C+2)-u)}\\
\alpha_{0}(u)      & := & \frac{1}{[C+2]_{q}}\{[C+1]_{q}\rho(u) - 1\}\\
\beta_{0}(u)       & := & \frac{1}{[C+2]_{q}}\{[2]_{q}\rho(u) - [C]_{q}\}\\
\gamma_{0}(u)      & := & \frac{1}{[C+2]_{q}}\{\rho(u) - [C+1]_{q}\}\\
\delta_{1}(u)      & := & \frac{1}{[C+2]_{q}}\{\rho(u)q^{-C-1} + q\}\\
\delta_{2}(u)      & := & \frac{1}{[C+2]_{q}}\{\rho(u)q^{C+1} + q^{-1}\}\\
\varepsilon_{1}(u) & := & \frac{\mu^{\ast}}{[C+2]_{q}}
                       \{\rho(u)q^{-\frac{C}{2}-1} + q^{\frac{C}{2}+1} \}\\
\varepsilon_{2}(u) & := & \frac{\mu}{[C+2]_{q}}
                       \{\rho(u)q^{\frac{C}{2}+1} + q^{-\frac{C}{2}-1} \}\\
\zeta_{1}(u)       & := & \frac{\kappa^{\ast}}{[C+2]_{q}}
                       \{\rho(u)q^{-\frac{C+1}{2}} + q^{\frac{C+1}{2}} \}\\
\zeta_{2}(u)       & := & \frac{\kappa}{[C+2]_{q}}
                       \{\rho(u)q^{\frac{C+1}{2}} + q^{-\frac{C+1}{2}} \}
\end{eqnarray}
\end{mathletters}
$f(u)$ and $g(u)$ in (\ref{RV4V4}) are defined by
\begin{mathletters}
\label{fg-def}
\begin{eqnarray}
f(u)   & = & \frac{\sinh(\eta(C+C^{\prime})\,+\,u)}
                  {\sinh(\eta(C+C^{\prime})\,-\,u)}\\
g(u)   & = & \frac{\sinh(\eta(C+C^{\prime}+2)\,-\,u)}
                  {\sinh(\eta(C+C^{\prime}+2)\,+\,u)}
\end{eqnarray}
\end{mathletters}
Using (\ref{$q$-bracket}) and the abbreviations 
\begin{eqnarray*}
\alpha      & = & [C+C^{\prime}]_{q}\\
\beta       & = & [C+C^{\prime}\,+\,1]_{q}\\
\gamma      & = & [C+C^{\prime}\,+\,2]_{q}\\
\varepsilon & = & [C^{\prime}]_{q}q^{\frac{C+C^{\prime}}{2}+1}
                  \,-\,[C]_{q}q^{-\frac{C+C^{\prime}}{2}-1}\\
\eta        & = & [C]_{q}q^{\frac{C+C^{\prime}}{2}+1}\,-\,
                  [C^{\prime}]_{q}q^{-\frac{C+C^{\prime}}{2}-1}\\
\end{eqnarray*}
the remaining coefficients of (\ref{RV4V4}) can be written as:
\begin{eqnarray*}
r_1 & = & \frac{\kappa^{\ast}\mu^{\ast}\kappa^{\prime}\mu^{\prime}}
                {\alpha\beta\gamma}
          (\gamma q^{C+C^{\prime}}f(u)+[2]_q\beta + 
                 \alpha q^{-C-C^{\prime}-2}g(u))\\
r_1^{\prime} & = & \frac{\kappa\mu{\kappa^{\prime}}^{\ast}
                          {\mu^{\prime}}^{\ast}}{\alpha\beta\gamma}
                    (\gamma q^{-C-C^{\prime}}f(u)+[2]_q\beta + 
                          \alpha q^{C+C^{\prime}+2}g(u))\\
r_2 & = & \frac{\kappa^{\ast}\kappa^{\prime}}{\alpha}
          \left(q^{\frac{C+C^{\prime}}{2}}f(u)+ 
                q^{-\frac{C+C^{\prime}}{2}}\right)\\
r_2^{\prime} & = & \frac{\kappa{\kappa^{\prime}}^{\ast}}{\alpha}
                   \left(q^{-\frac{C+C^{\prime}}{2}}f(u)+ 
                   q^{\frac{C+C^{\prime}}{2}}\right)\\
r_3 & = & \frac{\mu{\mu^{\prime}}^{\ast}}{\gamma}
          \left(q^{-\frac{C+C^{\prime}+2}{2}}+ 
                q^{\frac{C+C^{\prime}+2}{2}}g(u)\right)\\
r_3^{\prime} & = & \frac{\mu^{\ast}\mu^{\prime}}{\gamma}
                   \left(q^{\frac{C+C^{\prime}+2}{2}}+ 
                   q^{-\frac{C+C^{\prime}+2}{2}}g(u)\right)\\
r_4 & = & 1  + \frac{q^{-1}}{\alpha\beta\gamma}
          \big\{[C]_q\left([C^{\prime}]_q\gamma f(u)-
          [C+1]_{q}\beta\right)\\
    &   & \qquad + [C^{\prime}+1]_{q}\left([C+1]_q\alpha g(u)
                                   - [C^{\prime}]_{q}\beta \right)\big\}\\
r_4^{\prime} & = & 1  + \frac{q}{\alpha\beta\gamma}
          \big\{[C]_q\left([C^{\prime}]_q\gamma f(u)-
          [C+1]_{q}\beta\right)\\
    &   & \qquad + [C^{\prime}+1]_{q}\left([C+1]_q\alpha g(u)
                                   - [C^{\prime}]_{q}\beta \right)\big\}\\
r_5 & = & \frac{1}{\alpha\beta\gamma}
          \big([C]_q[C+1]_q\gamma f(u) - 
                [2]_q[C^{\prime}]_q[C+1]_q\beta\\ 
    &   & \qquad + [C^{\prime}][C^{\prime}+1]_q\alpha g(u)\big)\\
r_5^{\prime} & = & \frac{1}{\alpha\beta\gamma}
                   \big([C^{\prime}]_q[C^{\prime}+1]_q\gamma f(u) - 
                         [2]_q[C]_q[C^{\prime}+1]_q \beta\\ 
             &   & \qquad + [C][C+1]_q\alpha g(u)\big)\\
r_6 & = & \frac{\mu^{\ast}\kappa^{\prime}}{\alpha\beta\gamma}
          \big([C]_q\gamma q^{\frac{C+C^{\prime}+1}{2}}f(u) - 
                \beta\varepsilon q^{\frac{1}{2}}\\
    &   & \qquad - [C^{\prime}+1]_q\alpha
                q^{-\frac{C+C^{\prime}+1}{2}}g(u)\big)\\
r_6^{\prime} & = & \frac{\kappa{\mu^{\prime}}^{\ast}}{\alpha\beta\gamma}
                   \big([C^{\prime}]_q\gamma q^{-\frac{C+C^{\prime}+1}{2}}f(u)
                    + \beta\varepsilon q^{-\frac{1}{2}}\\ 
             &   & \qquad - [C+1]_q \alpha 
                   q^{\frac{C+C^{\prime}+1}{2}}g(u)\big)\\
r_7 & = & \frac{1}{\alpha}\left([C^{\prime}]_q - [C]_q f(u)\right)\\
r_7^{\prime} & = & \frac{1}{\alpha}\left([C]_q - 
                   [C^{\prime}]_q f(u)\right)\\
r_8 & = & \frac{\kappa^{\ast}\mu^{\prime}}{\alpha\beta\gamma}
          \big([C^{\prime}]_q\gamma q^{\frac{C+C^{\prime}+1}{2}}f(u)
                 - \beta\eta q^{\frac{1}{2}}\\
    &   & \qquad - [C+1]_q\alpha q^{-\frac{C+C^{\prime}+1}{2}}g(u)\big) \\
r_8^{\prime} & = & \frac{\mu{\kappa^{\prime}}^{\ast}}{\alpha\beta\gamma}
                   \big([C]_q\gamma q^{-\frac{C+C^{\prime}+1}{2}}f(u) + 
                \beta\eta q^{-\frac{1}{2}}\\
             &   & \qquad - [C^{\prime}+1]_q\alpha
                   q^{\frac{C+C^{\prime}+1}{2}}g(u)\big)\\
r_9 & = & \frac{1}{\gamma}\left([C^{\prime}+1]_q - [C+1]_q g(u)\right)\\
r_9^{\prime} & = & \frac{1}{\gamma}\left([C+1]_q - 
                                         [C^{\prime}+1]_q g(u)\right)\\
r_{10} & = & \frac{1}{\alpha\beta\gamma}
             \big\{[C]_{q}\left([C+1]_{q}\beta - [C^{\prime}]_q\gamma f(u)
                    \right)\\
       &   & \qquad + [C^{\prime}+1]_q\left([C^{\prime}]_{q}\beta
             -  [C+1]_q \alpha g(u)\right)\big\}\:.
\end{eqnarray*}
\section{Some details on ABA}
\label{AppB}
Applying the ansatz (\ref{NABA-es}) to the diagonal elements of 
$\hat{T}_{ij}(u)$ using (\ref{FCR1}) and (\ref{g-vac-act1}) yields \cite{KuRe}
\begin{mathletters}
\label{FCR3}
\begin{eqnarray}
\label{FCR3a}
&   & \left[\hat{T}_{11}(u)+\hat{T}_{22}(u)
      \right]|\lambda_{1},\ldots,\lambda_{M}|F>\\
\nonumber
& = & [\omega_{1}(u)]^{N}\prod_{i=1}^{M}\frac{1}{c(u-\lambda_{i})}
      [\hat{\tau}^{V_{2}}(M|u)]^{b_{1},\ldots,b_{M}}_{a_{1},\ldots,a_{M}}
      F^{a_{1},\ldots,a_{M}}\\
\nonumber
&   & \times \hat{C}_{b_{1}}(\lambda_{1})\cdots\hat{C}_{b_{M}}(\lambda_{M})
      \left|0\right>_{N}\\
\nonumber
& + & \sum_{k=1}^{M}\left[\check{\Lambda}^{(1,2)}_{k}
                          (u;\lambda_{1},\ldots,\lambda_{M})
      \right]^{b_{1},\ldots,b_{M}}_{a_{1},\ldots,a_{M}}
     F^{a_{1},\ldots,a_{M}}\\
\nonumber
&   & \times\hat{C}_{b_{k}}(u)\prod_{\stackrel{\s i=1}{\s i\neq k}}^{M}
      \hat{C}_{b_{i}}(\lambda_{i})\left|0\right>_{N}\:,
\end{eqnarray}
where $\hat{\tau}^{V_{2}}(M|u)$ is an inhomogeneous \emph{transfer-matrix} 
obtained according to (\ref{QISM1}) with $\delta^{(n)}=\gamma_{n}$ from 
(\ref{R6V}), and 
\begin{eqnarray}
\label{FCR3b}
&   & \hat{T}_{33}(u)|\lambda_{1},\ldots,\lambda_{M}|F>\\
\nonumber
& = & (-1)^{N}\prod_{i=1}^{M}\frac{1}{c(u-\lambda_{i})}
      |\lambda_{1},\ldots,\lambda_{M}|F>\\
\nonumber
& + & \sum_{k=1}^{M}\left[\check{\Lambda}^{(3)}_{k}
      (u;\lambda_{1},\ldots,\lambda_{M})
      \right]^{b_{1},\ldots,b_{M}}_{a_{1},\ldots,a_{M}}F^{a_{1},\ldots,a_{M}}\\
\nonumber
&   & \times\hat{C}_{b_{k}}(u)\prod_{\stackrel{\s i=1}{\s i\neq k}}^{M}
      \hat{C}_{b_{i}}(\lambda_{i})\left|0\right>_{N}
\end{eqnarray}
\end{mathletters}
The operators $\hat{C}_{i}(\lambda_{1})$ under the products in (\ref{FCR3}) 
are ordered with the index increasing from left to right factors. Note that 
only the first terms in equation (\ref{FCR3}) will contribute to the 
eigenvalue, while the following terms are \emph{unwanted}. Their coefficients 
$\check{\Lambda}_{k}$ are
\begin{mathletters}
\label{unwanted}
\begin{eqnarray}
\label{unwanted1}
&   & \left[\check{\Lambda}^{(1,2)}_{k}(u;\lambda_{1},\ldots,\lambda_{M})
      \right]^{b_{1},\ldots,b_{M}}_{a_{1},\ldots,a_{M}}\\
\nonumber
& = & -[\omega_{1}(\lambda_{k})]^{N}\frac{b(u-\lambda_{k})}{c(u-\lambda_{k})}
      \prod_{\stackrel{\s i=1}{i\neq k}}^{M}
             \frac{1}{c(\lambda_{k}-\lambda_{i})}
      \prod_{j=1}^{k-1}\frac{1}{d(\lambda_{j}-\lambda_{k})}\\
\nonumber
& \times & [\hat{L}_{c_{M}c_{M-1}}(\lambda_{k}-\lambda_{M})]^{b_{M}}_{a_{M}}
           [\hat{L}_{c_{M-1}c_{M-2}}(\lambda_{k}-\lambda_{M-1})
           ]^{b_{M-1}}_{a_{M-1}}\\
\nonumber
& \times & \cdots\times
           [\hat{L}_{c_{k+1}c_{k}}(\lambda_{k}-\lambda_{k+1})
           ]^{b_{k+1}}_{a_{k+1}}\\
\nonumber
& \times & \left(\prod_{l=1}^{k-1}\delta^{b_{l}}_{a_{l}}\right)
           \delta^{c_{k}}_{a_{k}}
           \left[\delta^{1}_{b_{k}}\delta^{c_{M}}_{1}
                +\delta^{2}_{b_{k}}\delta^{c_{M}}_{2}\right]
\end{eqnarray}
where $\hat{L}_{ij}(u)$ is an abbreviation for 
$[\hat{L}^{V_{2}}(n|u)]^{V_2}_{ij}$, derived from (\ref{R6V}) via 
(\ref{L-matrix}), and
\begin{eqnarray}
\label{unwanred2}
&   & \left[\check{\Lambda}^{(3)}_{k}(u;\lambda_{1},\ldots,\lambda_{M})
      \right]^{b_{1},\ldots,b_{M}}_{a_{1},\ldots,a_{M}}\\
\nonumber
& = & \frac{a(\lambda_{k}-u)}{c(\lambda_{k}-u)}(-1)^{N+M}
      \prod_{\stackrel{\s i=1}{i\neq k}}^{M}
      \frac{1}{c(\lambda_{i}-\lambda_{k})}
      \prod_{j=k+1}^{M}d(\lambda_{j}-\lambda_{k})\\
\nonumber
& \times & \left[\hat{S}_{k}(\lambda_{1},\ldots,\lambda_{k})
           \right]^{b_{1},\ldots,b_{k}}_{a_{1},\ldots,a_{k}}
           \left(\prod_{l=k+1}^{M}\delta^{b_{l}}_{a_{l}}\right)\:.
\end{eqnarray}
Here a $k$-particle $S$-matrix has been defined via \cite{KuRe}
\begin{eqnarray}
\label{unwanted3}
&   & \left[\hat{S}_{k}(\lambda_{1},\ldots,\lambda_{k})
      \right]^{b_{1},\ldots,b_{k}}_{a_{1},\ldots,a_{k}}\\
\nonumber
& = & \delta^{c_{1}}_{b_{k}}\delta^{c_{k}}_{a_{k}}
      \prod_{i=1}^{k-1}\,r_{b_{i}c_{i},a_{i},c_{i+1}}(\lambda_{i}-\lambda_{k}) 
\end{eqnarray}
\end{mathletters}
In (\ref{unwanted}) summation over repeated indices $c_{i}=1,2$ is
implicit. Applying the ansatz (\ref{NABA-es}) to (\ref{ev-equ})
forces the \emph{unwanted terms} in (\ref{FCR3}) to vanish. 
These equations can be transformed into \emph{6-vertex-type} eigenvalue 
equations (\ref{6V-ev-equ}) in section \ref{ABA} \cite{KuRe}.
\section{Derivation of Commutation Relations}
\label{AppC}
A few more details on the derivation of (\ref{FCR6}) are given:
In (\ref{FCR5c}) the term $\propto\hat{T}^{V_{3}}_{11}(u)
\hat{T}^{V^{\prime}_{4}}_{23}(v)$ acts non-trivially according to 
(\ref{g-vac-act1}) and (\ref{g-vac-act2}). It has to be eliminated by use of
\begin{eqnarray*}
\alpha_{0}(u-v)\hat{T}^{V_{3}}_{11}(u)\hat{T}^{V^{\prime}_{4}}_{23}(v) & & \\
  +\, \varepsilon_{1}(u-v)\hat{C}_{1}(u)\hat{T}^{V^{\prime}_{4}}_{33}(v)
& = & \rho(u-v)\hat{T}^{V^{\prime}_{4}}_{23}(v)\hat{T}^{V_{3}}_{11}(u)
\end{eqnarray*}
from (\ref{FCR4}), which results into (\ref{FCR6c}). Similarly (\ref{FCR5h}) 
can be handled, leading to (\ref{FCR6d}). In (\ref{FCR5d}) the term 
$\propto\hat{T}^{V_{3}}_{21}\hat{T}^{V^{\prime}_{4}}_{24}$ has to be 
eliminated via the relation
\begin{eqnarray*}
\alpha_{0}(u-v)\hat{T}^{V_{3}}_{21}(u)\hat{T}^{V^{\prime}_{4}}_{24}(v) & & \\
 +\,  \varepsilon_{1}(u-v)\hat{C}_{1}(u)\hat{T}^{V^{\prime}_{4}}_{44}(v)
& = & \alpha_{0}(u-v)\hat{T}^{V^{\prime}_{4}}_{24}(v)\hat{T}^{V_{3}}_{21}(u)\\
& + & \delta_{2}(u-v)\hat{T}^{V^{\prime}_{4}}_{23}(v)\hat{T}^{V_{3}}_{22}(u)\\
& + & \zeta_{2}(u-v)\hat{T}^{V^{\prime}_{4}}_{21}(v)\hat{T}^{V_{3}}_{32}(u)
\end{eqnarray*}
from (\ref{FCR4}). According to (\ref{g-vac-act1}) and (\ref{g-vac-act2})
the term $\propto\hat{T}^{V^{\prime}_{4}}_{43}(v)\hat{C}_{2}(u)$ also 
acts non-trivially on $\left|0\right>_{N}$. It has to be eliminated,
using the relations following relations from the set (\ref{FCR4}),
\begin{eqnarray*}
      \varepsilon_{2}(u-v)\hat{T}^{V_{3}}_{22}(u)
      \hat{T}^{V^{\prime}_{4}}_{23}(v) & & \\
  +\, \gamma_{0}(u-v)\hat{C}_{2}(u)\hat{T}^{V^{\prime}_{4}}_{43}(v)
& = & \delta_{1}(u-v)\hat{T}^{V^{\prime}_{4}}_{44}(v)\hat{C}_{1}(u)\\
& + & \alpha_{0}(u-v)\hat{T}^{V^{\prime}_{4}}_{43}(v)\hat{C}_{2}(u)\\
& + & \zeta_{2}(u-v)q\hat{T}^{V^{\prime}_{4}}_{41}(v)\hat{T}^{V_{3}}_{33}(u)
\end{eqnarray*}
and
\begin{eqnarray*}
      \alpha_{0}(u-v)\hat{T}^{V_{3}}_{22}(u)
      \hat{T}^{V^{\prime}_{4}}_{23}(v) & & \\
 +\,  \varepsilon_{1}(u-v)\hat{C}_{2}(u)\hat{T}^{V^{\prime}_{4}}_{43}(v)
& = & \delta_{1}(u-v)\hat{T}^{V^{\prime}_{4}}_{24}(v)\hat{T}^{V_{3}}_{21}(u)\\
& + & \alpha_{0}(u-v)\hat{T}^{V^{\prime}_{4}}_{23}(v)\hat{T}^{V_{3}}_{22}(u)\\
& + & \zeta_{2}(u-v)q\hat{T}^{V^{\prime}_{4}}_{21}(v)\hat{T}^{V_{3}}_{23}(u)
\end{eqnarray*}
Both relations have to be used in order to prevent the appearance of 
$\hat{T}^{V_{3}}_{22}(u)\hat{T}^{V^{\prime}_{4}}_{23}(v)$, 
also acting non-trivially on $\left|0\right>_{N}$, in the result (\ref{FCR6e}).
Applying the same procedure to (\ref{FCR5g}) leads to (\ref{FCR6f}).

\end{document}